\documentclass[aps,prd,nofootinbib,twocolumn,superscriptaddress,preprintnumbers,balancelastpage,longbibliography,nobibnotes]{revtex4-1}

\usepackage{xcolor}
\usepackage{rotating} % include this for rotate header
\usepackage{amsmath,amssymb,mathtools,bm}
\usepackage{graphicx, color, hepunits}
\usepackage[colorlinks=true
,urlcolor=blue
,anchorcolor=blue
,citecolor=blue
,filecolor=magenta
,linkcolor=red
,menucolor=blue
,linktocpage=true
,pdfproducer=medialab
,pdfa=true
]{hyperref}
\usepackage[utf8]{inputenc}
\usepackage[english]{babel}

\newcommand{\es}[2] {\begin{equation} \label{#1} \begin{split} #2 \end{split} \end{equation}}

\newcommand{\xx}{\mathbf{x}}
\newcommand{\vv}{\mathbf{v}}
\newcommand{\kk}{\mathbf{k}}
\newcommand{\vw}{v_{\omega}}
\newcommand{\FF}{\mathcal{F}}
\newcommand{\dd}{\mathbf{d}}
\newcommand{\mDM}{m_{\scriptscriptstyle {\rm DM}}}
\newcommand{\gagg}{g_{a \gamma \gamma}}
\newcommand{\rhoDM}{\rho_{\scriptscriptstyle {\rm DM}}}
\newcommand{\ldB}{\lambda_{\rm dB}}
\newcommand{\lc}{\lambda_c}
\newcommand{\bmt}{\bm{\theta}}
\newcommand{\bma}{\bm{\alpha}}
\newcommand{\bms}{\bm{\Sigma}}
\newcommand{\mcn}{\mathcal{N}}

\begin{document}

\title{Dark Matter Interferometry}

\author{Joshua W. Foster}
\affiliation{Leinweber Center for Theoretical Physics, Department of Physics, University of Michigan, Ann Arbor, MI 48109, U.S.A.}
\affiliation{Berkeley Center for Theoretical Physics, University of California, Berkeley, CA 94720, U.S.A.}
\affiliation{Theoretical Physics Group, Lawrence Berkeley National Laboratory, Berkeley, CA 94720, U.S.A.}

\author{Yonatan Kahn}
\affiliation{Department of Physics, University of Illinois at Urbana-Champaign, Urbana, IL 61801, U.S.A.}
\affiliation{Illinois Center for Advanced Studies of the Universe, University of Illinois at Urbana-Champaign, Urbana, IL 61801, U.S.A.}

\author{Rachel Nguyen}
\affiliation{Department of Physics, University of Illinois at Urbana-Champaign, Urbana, IL 61801, U.S.A.}
\affiliation{Illinois Center for Advanced Studies of the Universe, University of Illinois at Urbana-Champaign, Urbana, IL 61801, U.S.A.}

\author{Nicholas L. Rodd}
\affiliation{Berkeley Center for Theoretical Physics, University of California, Berkeley, CA 94720, U.S.A.}
\affiliation{Theoretical Physics Group, Lawrence Berkeley National Laboratory, Berkeley, CA 94720, U.S.A.}

\author{Benjamin R. Safdi}
\affiliation{Leinweber Center for Theoretical Physics, Department of Physics, University of Michigan, Ann Arbor, MI 48109, U.S.A.}
\affiliation{Berkeley Center for Theoretical Physics, University of California, Berkeley, CA 94720, U.S.A.}
\affiliation{Theoretical Physics Group, Lawrence Berkeley National Laboratory, Berkeley, CA 94720, U.S.A.}

\begin{abstract}
The next generation of ultralight dark matter (DM) direct detection experiments, which could confirm sub-eV bosons as the dominant source of DM, will feature multiple detectors operating at various terrestrial locations.
As a result of the wave-like nature of ultralight DM, spatially separated detectors will each measure a unique DM phase.
When the separation between experiments is comparable to the DM coherence length, the spatially-varying phase contains information beyond that which is accessible at a single detector.
We introduce a formalism to extract this information, which performs interferometry directly on the DM wave.
In particular, we develop a likelihood-based framework that combines data from multiple experiments to constrain directional information about the DM phase space distribution.  We show that the signal in multiple detectors is subject to a daily modulation effect unique to wave-like DM.
Leveraging daily modulation, we illustrate that within days of an initial discovery multiple detectors acting in unison could localize directional parameters of the DM velocity distribution such as the direction of the solar velocity to sub-degree accuracy, or the direction of a putative cold DM stream to the sub-arcminute level.
We outline how to optimize the locations of multiple detectors with either resonant cavity (such as ADMX or HAYSTAC) or quasistatic (such as ABRACADABRA or DM-Radio) readouts to have maximal sensitivity to the full 3-dimensional DM velocity distribution.
\end{abstract}
\maketitle

%%%%%%%%%%%%%%%%%
\section{Introduction}
%%%%%%%%%%%%%%%%%

Cold, bosonic dark matter (DM) candidates with masses much smaller than the eV scale have macroscopic occupation numbers and may be described in the solar vicinity by classical fields.
Two well-studied DM candidates in this category, which we broadly refer to as wave-like DM, are axions~\cite{Preskill:1982cy,Abbott:1982af,Dine:1982ah,Peccei:1977hh,Peccei:1977ur,Weinberg:1977ma,Wilczek:1977pj,Dine:1981rt,Zhitnitsky:1980tq,Kim:1979if,Shifman:1979if} and dark photons~\cite{Nelson:2011sf,Ahlers:2007rd,Jaeckel:2007ch}.
Wave-like DM candidates require distinctive  experimental techniques for discovery that take advantage of their spatial and temporal coherence (see, {\it e.g.},~\cite{Battaglieri:2017aum}).
The spatial coherence length of the DM waves, $\lc$, and the coherence time $\tau$ are given by\footnote{We briefly review both concepts in App.~\ref{app:coherence}.}~\cite{Foster:2017hbq}
\es{eq:lcdef}{
\lc \sim \frac{1}{\mDM v_0} \,, \qquad \tau \sim \frac{1}{\mDM \bar{v}\, v_0} \,,
}
where $v_0$ parameterizes the DM velocity dispersion, $\bar{v}$ is the mean velocity, and $\mDM$ is the DM mass.
In the solar neighborhood we expect $\bar{v} \sim v_0 \sim 10^{-3}$ for the bulk of the DM, in natural units, such that the coherence length is around $10^3$ times the Compton wavelength, and the coherence time is around $10^6$ times the oscillation period for the DM wave.
In this work we show that multiple phase-sensitive wave-like DM detectors separated by distances of order $\lc$ may join their data -- through a process we refer to as ``DM interferometry" -- to measure properties of the DM phase-space distribution that are inaccessible  to single experiments operating in isolation.

Many axion and dark-photon detection strategies already leverage the axion coherence \emph{time} as a ``quality factor" that amplifies the DM signal in the experiment.
For example, axion haloscopes~\cite{Sikivie:1983ip,Du:2018uak,McAllister:2017lkb,Brubaker:2017rna,Backes:2020ajv,Lee:2020cfj} use a resonant cavity with a strong static magnetic field to convert axion DM into electromagnetic cavity modes, which build up coherently over the DM coherence time; in this setup the DM Compton wavelength is of order the size of the experiment.
Experiments operating in the quasistatic regime (where the DM Compton wavelength is much larger than the experiment) -- including searches for the axion-photon coupling~\cite{Kahn:2016aff,Ouellet:2018beu,Ouellet:2019tlz,Zhong:2018rsr,Gramolin:2018qnt,Gramolin:2020ict}, axion interactions with nuclear spins~\cite{Budker:2013hfa}, or dark photons~\cite{Chaudhuri:2014dla} -- aim to detect a time-varying magnetic flux through a pickup loop, which can build up coherently in a lumped-element circuit~\cite{Chaudhuri:2019ntz,Chaudhuri:2018rqn}.

\begin{figure*}[t]
\centering	
\includegraphics[width=.47\textwidth]{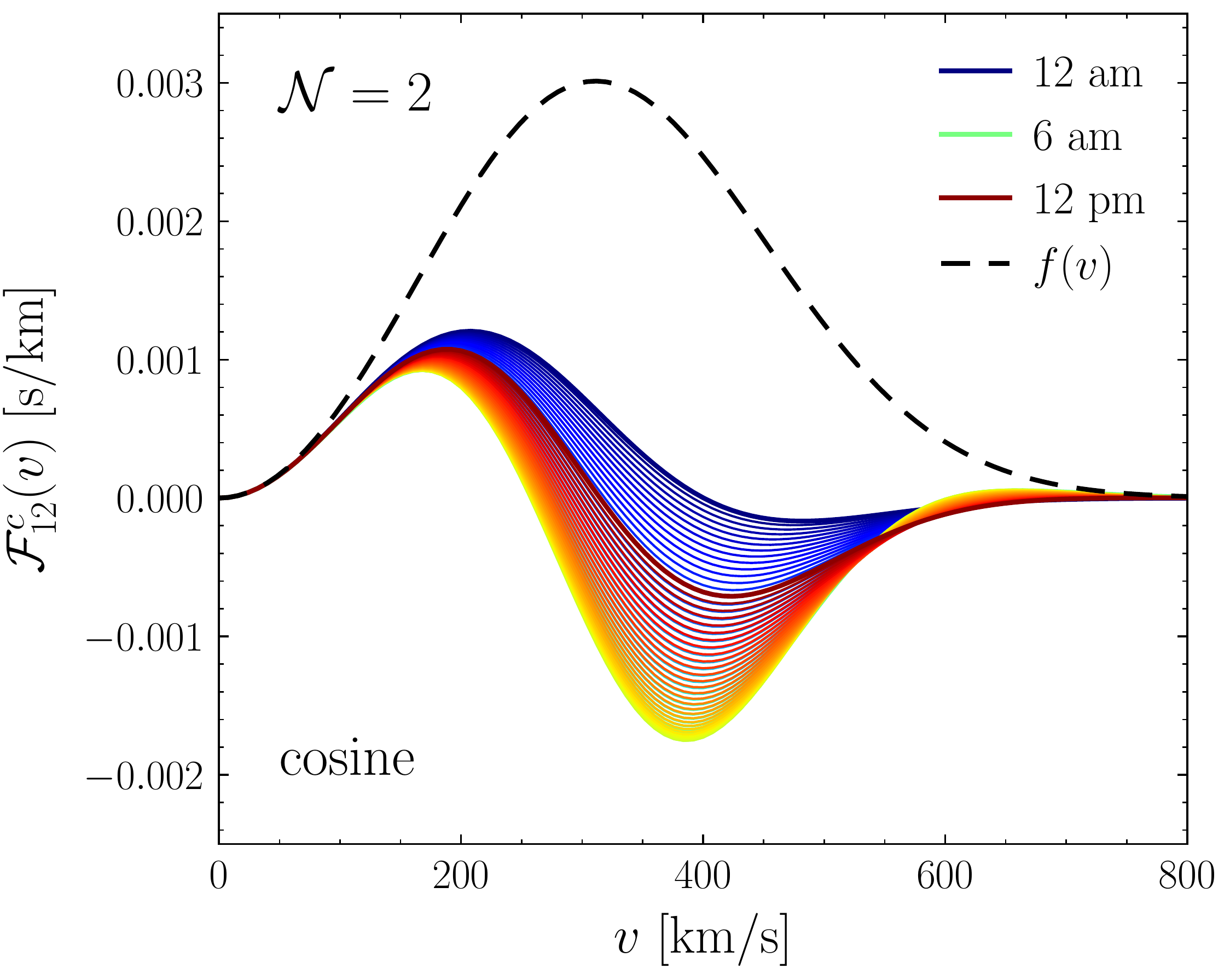}\hspace{0.2cm}
\includegraphics[width=.47\textwidth]{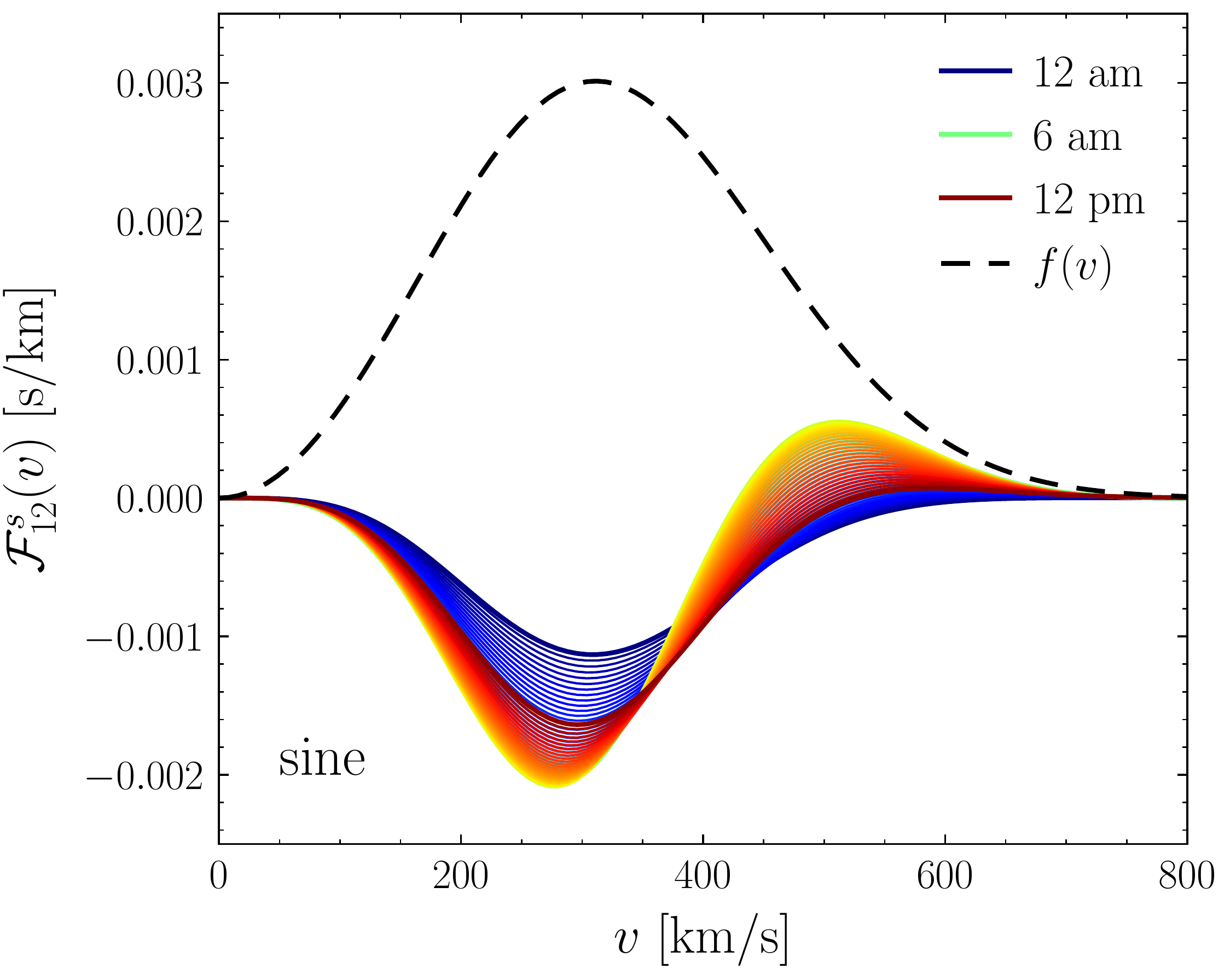}
\vspace{-0.3cm}
\caption{The imprint of DM interferometry.
A single wave-like DM experiment is sensitive to the DM speed distribution $f(v)$.
Two detectors separated by a vector $\xx_{12}$, however, are sensitive to the speed distribution modulated by the $\kk \cdot \xx_{12}$ phase of the DM wave, replacing $f(v)$ with functions $\FF^{c,s}_{12}(v)$ as defined in \eqref{eq:FcsIntro}.
As the figures demonstrate, the modified speed distributions exhibit daily modulation and carry additional information about the velocity distribution $f(\vv)$ that would be invisible to a single detector.
For this example we take $\mDM = 25.2~\mu{\rm eV}$~\cite{Buschmann:2019icd}, near the window where the HAYSTAC collaboration is searching for axion DM.
Taking the Standard Halo Model ansatz for $f(\vv)$ in \eqref{eq:SHM_full}, we place one detector at a latitude and longitude of $(41^{\circ}~{\rm N},\,73^{\circ}~{\rm W})$, and a second instrument $\sim 20~{\rm m}$ to the North, corresponding to $d \sim 2 \lc$.
A curve is shown for every ten minutes starting from midnight on January 1st of 2020.
Note that as $\FF_{12}^{c,s}(v)$ are functions of $\mDM d$, qualitatively similar effects exist for {\it e.g.} $\mDM \sim 10^{-9} \ {\rm eV}$, in the mass range probed by ABRACADABRA and DM-Radio, for $d \sim 500$ km.
}
\label{fig:exampleF}
\end{figure*}

In this paper, we explore the phenomenology of spatial coherence for wave-like DM by exploiting spatially-separated detectors that probe the same DM field.
It is straightforward to understand why multiple detectors offer unique insights for wave-like DM.
Generically, the wave-like DM field may be written as $a(\xx,t) = a_0 \cos(\omega t - \kk \cdot \xx + \phi)$, where $\omega$ is the oscillation frequency, $\kk$ is the wave vector, $\phi$ is a random phase, and $a_0$ is the amplitude.\footnote{Vector DM also has a polarization component with nontrivial coherence properties, but in this work we focus only on the amplitude, as appropriate for scalar or pseudoscalar DM.}
If the DM wave is traveling in the direction ${\bf \hat k}$ with speed $v \ll 1$, then  $\omega \approx \mDM(1+v^2/2)$ and $\kk \approx \mDM v {\bf \hat k}$. For a single detector we may always choose coordinates such that $\xx = 0$.
This means that a single detector is only sensitive to the speed through $\omega$ and is not sensitive to the direction of the DM velocity.\footnote{Exceptions would be experiments that make use of $\nabla a$, but such signals are suppressed by $v \sim 10^{-3}$ relative to experiments that are also sensitive to $\partial_t a$.
Experiments only sensitive to the speed distribution may also detect annual modulation signals through shifts in the DM speed~\cite{OHare:2017yze}, though these are typically quite small because the Earth's speed relative to the Sun is small compared to the solar speed relative to the Galactic Center.}
By contrast, two experiments located at positions $\xx_1$ and  $\xx_2$ will be sensitive to phase factors $\kk \cdot \xx_1$ and $\kk \cdot \xx_2$.
Only one of these can be removed by a coordinate choice, leaving a residual $\kk \cdot \xx_{12}$, with $\xx_{12} = \xx_1-\xx_2$, which manifestly probes the velocity rather than the speed.
The interferometry proposed in this work is directly at the level of the DM field: the effect arises due to the phase difference wave-like DM exhibits between spatially separated points.\footnote{This is conceptually distinct from the interferometry proposed in Refs.~\cite{DeRocco:2018jwe,Obata:2018vvr,Liu:2018icu}, where the interference results from a phase shift developed by electromagnetic fields as they propagate through axion DM.}
Indeed, due to the nonzero velocity dispersion $v_0$, DM waves are coherent up to distances of order $\lc$; as we will show, phase-sensitive data combined from two experiments exhibits maximal modulation when $d \equiv |\xx_{12}| \sim \lc$, or
\es{eq:dlDB}{
d \sim \frac{1}{\mDM v_0} = 270 \ {\rm km} \left(\frac{220 \ {\rm km/s}}{v_0}\right) \left(\frac{10^{-9} \ {\rm eV}}{\mDM}\right).
}
As we will demonstrate, this opens up striking new signatures, such as a unique daily modulation signal applicable only to wave-like DM with multiple detectors, because the direction of $\xx_{12}$ rotates over a sidereal day with respect to the DM field.\footnote{Several experimental proposals have noted or exploited sensitivity to the coherence length, see {\it e.g.}~\cite{TheMADMAXWorkingGroup:2016hpc,Millar:2016cjp,Knirck:2018knd,Millar:2017eoc,Lawson:2019brd,Knirck:2019eug,Arvanitaki:2017nhi}, but here we focus specifically on combining data between different experiments.}

The main result of this paper is that interference effects between a pair of detectors separated by a distance $\xx_{12}$ are fully characterized by the modified speed distributions
\es{eq:FcsIntro}{
\FF^c_{12}(v) =  \int d^3 \vv f(\vv) \cos(\mDM \vv \cdot \xx_{12}) \delta[ |\vv|-v], \\
\FF^s_{12}(v) =  \int d^3 \vv f(\vv) \sin(\mDM \vv \cdot \xx_{12}) \delta[ |\vv|-v],
}
with $f(\vv)$ the DM velocity distribution. Examples of these distributions at various times throughout the day are shown in Fig.~\ref{fig:exampleF} for optimally-separated detectors. 
If the goal is simply to enhance the total signal reach, we should maximize the constructive interference and take $d \ll \lc$, in which case $\FF_{12}^s(v) = 0$ and $\FF_{12}^c(v) = f(v)$, with $f(v)$ the DM speed distribution.
The observation that there is an enhanced sensitivity for an array of detectors located within the DM coherence length has been made previously in Ref.~\cite{Derevianko:2016vpm}; this is also the basis for the multiplexed cavity setup proposed by ADMX~\cite{Shokair:2014rna}.
However, if the goal is to extract information about the full 3-dimensional DM phase space distribution, which encodes {\it e.g.} the boost of the Solar System with respect to the Galactic Center as well as possible DM substructure (including the Sagittarius stream~\cite{10.1093/mnras/275.2.429} and the {\it Gaia} Sausage~\cite{10.1093/mnras/sty982,2018Natur.563...85H}), we should take $d \sim \lc$ as in~\eqref{eq:dlDB}.
Ref.~\cite{Derevianko:2016vpm} points out the possibility of observing this daily modulation effect for experiments separated by distances of order $\lc$; here we extend this analysis by focusing on constraining directional parameters in the phase space distribution.
We will show that the sensitivity to the phase space information that may be extracted from multiple detectors is comparable to the sensitivity to the initial discovery, since the interference effects have an $\mathcal{O}(1)$ effect on the data when $d \sim \lc$.
As such, in principle these unique signatures could be used to immediately verify a putative axion signal. 
More optimistically, DM interferometry would allow for the detailed mapping of the local DM phase space distribution after an initial detection.

For concreteness, we focus in this work on the case of axion DM coupled to electromagnetic signals, but our results would apply equally well to scalar and vector DM as long as the readout is proportional to the DM field.
Similarly, for simplicity we will present most results for the case of two experiments, but our formalism holds for any number $\mcn \geq 2$ of experiments, and we will provide our key results for a general $\mcn$ also.
Our results also apply equally well to resonant-type experiments and to broadband-type experiments (such as ABRACADABRA-10 cm~\cite{Ouellet:2018beu,Ouellet:2019tlz}), so long as the resonant experiments are able to preserve the phase of the data, as opposed to {\it e.g.} recording the power directly.
One advantage of resonant experiments for wave-like DM, in addition to generically having enhanced sensitivity~\cite{Chaudhuri:2019ntz,Chaudhuri:2018rqn}, is that putative signal candidates may immediately yield detailed and high-significance studies, since the signal-to-noise ratio rapidly grows with measurement time when frequency-scanning is no longer necessary.

We organize the remaining discussion as follows.
In Sec.~\ref{sec:stats}, we sketch a derivation for the statistics of the correlated Fourier-transformed data from multiple experiments.  A more extensive derivation and discussion is presented in App.~\ref{app:derivation}, with some useful orthogonality relations summarized in App.~\ref{app:orthogonality}.
In Sec.~\ref{sec:likelihood}, we construct a likelihood function for the axion signal as observed at $\mcn$ experiments, following the formalism of~\cite{Foster:2017hbq}; a practical data-stacking procedure is outlined in App.~\ref{app:stack}.
In Sec.~\ref{sec:parameters} we use several simplified toy examples to illustrate analytic estimates of uncertainties on parameters of the velocity distribution using the Asimov data set -- a technique where the asymptotic properties of the data are assumed in order to replace Monte Carlo simulations with analytic estimates (see Sec. \ref{sec:likelihood}) -- and demonstrate that uncertainties on directional parameters in several simple examples with two detectors are minimized for $d \sim 2\lc$. We also highlight the important distinction between the de Broglie wavelength and the coherence length for cold but boosted DM substructure. Furthermore, we show that there is a rotational symmetry of the likelihood which can lead to degenerate best-fit parameters for $\mathcal{N} = 2$ experiments.
In Sec.~\ref{sec:daily}, we extend the likelihood analysis to include daily modulation from the changing detector orientation throughout the day.
We also perform analyses of simulated data sets to demonstrate how the likelihood may be implemented in practice to constrain the morphology of the DM phase-space distribution. Using the realistic examples of the Standard Halo Model (SHM) velocity distribution and the Sagittarius Stream, we show how daily modulation breaks the symmetry discussed in Sec. \ref{sec:parameters} and use this to perform parameter estimation using the effect.
We conclude in Sec.~\ref{sec:conclusions} with some practical implications for current and upcoming axion experiments.
In App.~\ref{app:coherence} we provide a brief review of the coherence length and time.

%%%%%%%%%%%%%%%%%
\section{The Statistics of Multiple Detectors}
\label{sec:stats}
%%%%%%%%%%%%%%%%%

In this section we describe the statistics of an axion DM signal collected by two or more spatially-separated detectors.
In particular, while we expect background sources to be generally uncorrelated between detectors, the axion will induce non-trivial cross-correlations indicative of DM interferometry.
These correlations will be the source of the additional information available to two or more experiments that we will extract using a likelihood formalism introduced in Sec.~\ref{sec:likelihood}.

We imagine that a given detector, located at a position $\xx$, is sensitive to the axion through a time-varying signal $\Phi$ proportional to the axion field,
\es{eq:Phiexp}{
\Phi(\xx, t) = m_a \kappa_i \, a(\xx,t).
}
The flux $\Phi$ is generated by the axion effective current, $\mathbf{J}_a \sim \partial_t a \sim m_a a$, which is the origin of $m_a$ in the expression.
Accordingly, $\Phi \sim \kappa_i \mathbf{J}_a$, revealing $\kappa_i$ as characterizing the individual experimental response to the axion field.
In the notation of Ref.~\cite{Foster:2017hbq}, we take $\kappa_i = \sqrt{A_i/\rhoDM}$.
The dimensionful constant $A_i$ is characteristic of the individual experimental response to the axion field; for instance, in the case of ABRACADABRA~\cite{Kahn:2016aff,Ouellet:2018beu,Ouellet:2019tlz}, the magnetic flux induced in the pickup loop at the center of the detector is related to the axion field by $A = \rhoDM \gagg^2 B_0^2 V_B^2$, where $\gagg$ is the axion-photon coupling, $B_0$ is the toroidal magnetic field strength, and $V_B$ is an effective magnetic field volume associated with the detector.
In addition to the experimental factors, $A_i$ has been defined to include $\rhoDM \gagg^2$, which determines the mean power in the axion field.
We assume for simplicity that the detector response $A_i$ is purely real and does not include phase delays. Similar expressions are available for other detectors~\cite{Foster:2017hbq}.
For our discussion, all that is required is a measurement linear in the axion field, in order to ensure direct access to the axion phase.  Measurements intrinsically proportional to $a^2$, such as the power in the cavity of an axion haloscope, cannot be directly ported to our formalism.
Nevertheless, interferometry can still be performed by these resonant cavity experiments, as long as the phase information is extracted.
This may be achieved for example by reading out electromagnetic signals with phase-sensitive amplifiers ({\it e.g.},~\cite{Malnou:2018dxn,Backes:2020ajv}).

Ultimately, we envision a set of measurements $\Phi_i$ of the same axion field, made by $\mcn$ detectors at different spatial locations $\xx_i$.
The correlations between these data sets will arise due to the statistics of the underlying axion field, as we will describe in the following subsections, leaving the full derivation to App.~\ref{app:derivation}.

%%%%%%%%%%%%%%%%%
\subsection{Construction of the Axion Field}
%%%%%%%%%%%%%%%%%

It is useful to recall the underlying statistics in the axion field that result from its finite velocity dispersion and wave-like nature.
In~\cite{Foster:2017hbq} it was shown that we may represent the axion field as seen by a single detector as 
\es{eq:axion-construction-0}{
a(t) &= \frac{\sqrt{\rhoDM}}{m_a}\sum_{j} \alpha_j \sqrt{f(v_j) \Delta v}\cos \left[\omega_j t  + \phi_{j} \right].
} 
Here, the sum over $j$ indicates a sum over subsets of particles with speeds in the interval $v$ to $v + \Delta v$. The phase is controlled by $\omega_j = m_a \left( 1 + v_j^2/2 \right)$ and a random contribution $\phi_j \in [0, 2 \pi)$, and further $f(v)$ is the DM speed distribution in the laboratory frame. In Ref.~\cite{Foster:2017hbq} the continuum limit for speeds $\Delta v \to 0$ is taken; below (and in App.~\ref{app:derivation}) we will generalize to the continuum limit for velocities. 
In addition to the random phase, the random nature of the axion field is captured in the random variate $\alpha_j$ drawn from the Rayleigh distribution $p[\alpha] = \alpha\, e^{-\alpha^2/2}$.

While \eqref{eq:axion-construction-0} represents the axion field constructed from the discretized frequency modes specified by the local DM velocity distribution, a more fundamental approach can be understood by considering the local DM field made up of $N_a$ axion particles (or wave packets), as detailed in~\cite{Foster:2017hbq}.
The enormous occupation numbers characteristic of wave-like DM will then allow us to eventually convert this sum to an integral by taking the $N_a \to \infty$ limit; in detail, we should have $n_{\scriptscriptstyle {\rm DM}} \ldB^3 \gg 1$, where $n_{\scriptscriptstyle {\rm DM}}$ is the DM number density, which is satisfied locally for $m_a \ll 1~{\rm eV}$.
We note that the above construction also assumes DM is a non-interacting wave, which means that self-interactions should be negligible.

The axion field described in~\eqref{eq:axion-construction-0} is appropriate for a single detector, but to reveal the effects of DM interferometry we need to extend the description to include the spatial dependence of the DM wave.
In particular, the phase will also include a contribution $\kk \cdot \xx$, with $\kk = m_a \vv$ for a non-relativistic wave.
As $\kk$ depends on the velocity, and not speed, we need to extend the above sum to three independent components, $\vv_{abc} = v_a \hat{\xx} + v_b \hat{\bf y} + v_c \hat{\bf z}$, where the indexes $a,b,c$ are integers.
We may then write
\es{eq:axionfield}{
a(\xx, t) &= \frac{\sqrt{\rhoDM}}{m_a}\sum_{abc} \alpha_{abc} \sqrt{f(\vv_{abc}) (\Delta v)^3}
\\
&\times \cos \left[\omega_{abc} t - \kk_{abc} \cdot \xx + \phi_{abc} \right] ,
}
where $\omega_{abc}$ depends on $v_{abc} = |\vv_{abc}|$, and $\alpha_{abc}$ and $\phi_{abc}$ are Rayleigh and uniform random variables, respectively, as in~\eqref{eq:axion-construction-0}. Here, $(\Delta v)^3$ is a discretization of the 3-dimensional velocity, generalizing $\Delta v$ for speeds; we will take the continuum limit in App.~\ref{app:derivation}.

In \eqref{eq:axionfield} we have written the axion field in a convenient form for revealing DM interferometry.
To reiterate the point, if we measure the axion field at a single location, we can always choose our coordinates such that $\xx = 0$.
In this case, the velocity information in $\kk$ is lost, and we are now only sensitive to the speed $v = |\vv|$ through $\omega$.
This collapses $f(\vv) \to f(v)$, and~\eqref{eq:axionfield} to~\eqref{eq:axion-construction-0}; information about the phase space is lost.
However, if we measure the axion field at two locations, an irreducible $\kk$ dependence remains, and the full velocity information is imprinted in the multi-detector covariance matrix.

We implicitly assume throughout this work that the non-interacting plane-wave superposition in~\eqref{eq:axionfield} applies for all $\xx$.
Corrections to this picture should arise from {\it e.g.} the gravitational field of the Earth, which would slightly bend the DM trajectories between detectors.
However, the DM velocities we consider in this work are much larger than the Earth's escape velocity, and also the detector separations are typically much smaller than the radius of the Earth, so we are justified in neglecting this effect.

%%%%%%%%%%%%%%%%%
\subsection{The Multi-Detector Covariance Matrix}
%%%%%%%%%%%%%%%%%

We will now outline how the statistics of the axion field, as described above, lead to a non-trivial covariance matrix in the data collected by $\mcn$ experiments.
In this section we will simply state the key results, leaving a derivation to App.~\ref{app:derivation}.

We begin by considering the minimal case of $\mcn=1$.
Suppose that a single experiment takes a time-series of $N$ measurements $\{\Phi_n(\xx) = \Phi(\xx, n \Delta t)\}$, with $n=0,1,\ldots,N-1$, collected over a time $T$, so that $\Delta t = T/N$.
In order to isolate a signal oscillating at a particular frequency, as we expect the axion to do, we calculate the discrete Fourier transform
\es{eq:DFT}{
\Phi_k(\xx) = \sum_{n=0}^{N-1}\Phi_n(\xx)\, e^{-i2 \pi k n /N} \,.
}
The transform is indexed by an integer $k=0,1,\ldots,N-1$, which is related to the angular frequency, $\omega = 2 \pi k/T$.
We will switch back and forth between talking about frequency $\omega$ and wave-number $k$ as convenient.
It is convenient to partition the Fourier transform into appropriately normalized real and imaginary parts as follows,
\es{eq:re-im}{
R_k({\bf x}) &= \frac{\Delta t}{\sqrt{T}} {\rm Re}\left[ \Phi_k({\bf x}) \right] , \\
I_k({\bf x}) &= \frac{\Delta t}{\sqrt{T}} {\rm Im}\left[ \Phi_k({\bf x}) \right] .
}
We can then write the power spectral density (PSD) as,
\es{eq:PSDdef}{
S^k_{\Phi \Phi} = \frac{(\Delta t)^2}{T} |\Phi_k(\xx)|^2 = R_k^2({\bf x}) +  I_k^2({\bf x}) .
}
We will assume throughout that $T$ is long enough such that the signal is sufficiently well resolved, {\it i.e.} the bandwidth of the Fourier transform $2 \pi / T$ is much smaller than the width of the signal in frequency space. 
When specifying Fourier components by frequency as opposed to wave-number we use notation as in $S_{\Phi\Phi}^k({\bf x}) \to S_{\Phi \Phi}({\bf x}, \omega)$.

As shown in~\cite{Foster:2017hbq}, both $R({\bf x},\omega)$ and $I({\bf x},\omega)$ are normally distributed with zero mean and variance given by
\es{eq:single-var}{
\langle R^2({\bf x},\omega) \rangle = \langle I^2({\bf x},\omega) \rangle = \frac{A}{2}  \frac{\pi f(\vw)}{m_a \vw} \,,
}
where we have defined $\vw =\sqrt{2\omega/m_a -2}$ as the axion velocity corresponding to a frequency $\omega$, and the speed distribution is defined as
\begin{equation}
     f(v) = \int d^3\mathbf{v} f(\mathbf{v}) \delta(v - |\mathbf{v}|)
\end{equation}

This implies, for example, that $S_{\Phi\Phi}(\xx)$ is an exponentially distributed quantity, with mean
\es{eq:singlePSD}{
\langle S_{\Phi\Phi} (\omega) \rangle = A \frac{\pi f(\vw)}{m_a \vw}\,.
}
We can understand the velocity dependence by looking back to \eqref{eq:axion-construction-0}; the signal measured by a detector $\Phi$ is proportional to the time-dependent axion axion field, and since $a$ is proportional to $\sqrt{f(v)}$, we obtain a power spectrum $S_{\Phi \Phi}$ proportional to $f(v)$.

In any real experiment there will also be background.
However, as long as the background is normally distributed in the time domain -- as expected for, for example, thermal noise, SQUID flux noise, or Josephson parametric amplifier noise -- then both $R({\bf x},\omega)$ and $I({\bf x},\omega)$ remain normally distributed but with variance 
\es{eq:lambda_omega}{
\langle R^2({\bf x},\omega) \rangle = \langle I^2({\bf x},\omega) \rangle =\frac{A}{2}  \frac{\pi f(\vw)}{m_a \vw} + \frac{\lambda_B(\omega)}{2} \,,
}
where $\lambda_B(\omega)$ encapsulates the variance of the potentially frequency-dependent noise from the background sources only.

Note that $R_k(\xx)$ and $I_k(\xx)$ are uncorrelated; in particular, the $2 \times 2$ covariance matrix for these two quantities is simply
\es{eq:SigmaN1}{
\bms_k = \left (\frac{A}{2}  \frac{\pi f(\vw)}{m_a \vw} + \frac{\lambda_B(\omega)}{2} \right)
\begin{bmatrix}
 1 &  0  \\
 0 &  1  \\
\end{bmatrix}.
}
This implies that for a single detector, all information about the signal is contained in the PSD $S_{\Phi\Phi}(\xx)$.
Further, as shown in~\eqref{eq:SigmaN1}, the location ${\bf x}$ never enters for $\mcn=1$.
Even if we chose our coordinates such that $\kk \cdot \xx \neq 0$, the overall phase remains unphysical as it would vanish when computing the modulus squared in~\eqref{eq:PSDdef}.

Now let us extend the discussion to the case of interest: data collected by $\mcn$ experiments at positions $\xx_i$, with $i = 1,2,\ldots, \mcn$.
For each data set, we calculate the real and imaginary parts of the Fourier transform as above.
The information collected by all detectors can then be organized into the following $2\mcn$ dimensional data vector,
\es{eq:data_vector}{
\dd_k = \left[ R_k( \xx_1),I_k( \xx_1), \ldots, R_k( \xx_\mcn),R_k( \xx_\mcn) \right]^{T} .
}
Correlations between the real and imaginary part for any given detector will be identical to the $\mcn=1$ case discussed above.
However, DM interferometry will reveal itself through non-trivial correlations amongst the different detectors.\footnote{Our analysis assumes that the experiments have identical timestamps on the data, or equivalently that the relative phase of the signal at each experiment is precisely known. Of course, this is not exactly true and in general there will be an additional contribution to the phase of $\Phi$ in \eqref{eq:DFT} of the form $\omega \Delta \tau$, where $\Delta \tau$ is the timing error.
As long as $ \Delta \tau \ll |\xx_{ij}| v$, this contribution can be safely neglected. 
For two detectors with $|\xx_{12}| \sim$ 50 m and $v \sim 200$ km/s, this implies $\Delta \tau \ll  10^{-10}$ s.
Typical atomic clocks have timing error of $10^{-9} \ {\rm s}$/day, so the required $\Delta \tau$ can be achieved by synchronizing the two experiments to an atomic clock over data-taking intervals of about 2.5 hours, which is sufficient for the daily modulation analysis in Sec.~\ref{sec:daily}.}
Indeed, as justified in App.~\ref{app:derivation}, $\dd_k$ will be a $2 \mcn$-dimensional Gaussian random variable with zero mean and a symmetric $(2 \mcn \times 2 \mcn)$-dimensional covariance matrix given by
\begin{widetext}
\es{eq:fullcovariance}{
\bms_k = &
\begin{bmatrix}
\langle R_k(\xx_1) R_k(\xx_1) \rangle
& \langle R_k(\xx_1) I_k(\xx_1) \rangle
& \langle R_k(\xx_1) R_k(\xx_2) \rangle
& \ldots
& \langle R_k(\xx_1) I_k(\xx_{\mcn}) \rangle \\
\langle I_k(\xx_1) R_k(\xx_1) \rangle
& \langle I_k(\xx_1) I_k(\xx_1) \rangle
& \langle I_k(\xx_1) R_k(\xx_2) \rangle
& \ldots
& \langle I_k(\xx_1) I_k(\xx_{\mcn}) \rangle \\
\vdots & & \ddots \\
\langle I_k(\xx_{\mcn}) R_k(\xx_1) \rangle
& \langle I_k(\xx_{\mcn}) I_k(\xx_1) \rangle
& \langle I_k(\xx_{\mcn}) R_k(\xx_2) \rangle
& \ldots
& \langle I_k(\xx_{\mcn}) I_k(\xx_{\mcn}) \rangle
\end{bmatrix}, \\
\langle R_k(\xx_i) R_k(\xx_j) \rangle 
=\,& \langle I_k(\xx_i) I_k(\xx_j) \rangle 
= \frac{1}{2} \left[ c_{ij}(\omega)+\delta_{ij} \lambda_{B,i} (\omega) \right], \\
\langle R_k(\xx_i) I_k(\xx_j) \rangle 
=\,& -\langle I_k(\xx_i) R_k(\xx_j) \rangle 
= \frac{1}{2} s_{ij}(\omega).
}
\end{widetext}
Here $\lambda_{B,i}(\omega)$ is the background observed by the $i^{\rm th}$ experiment, and its contribution is purely diagonal.
The axion signal, however, induces off-diagonal correlations, which we quantify in terms of
\es{eq:c-s}{
c_{ij}(\omega) &= \frac{\pi \sqrt{A_i A_j}}{m_a \vw} \FF_{ij}^c(\vw) , \\
s_{ij}(\omega) &= \frac{\pi \sqrt{A_i A_j}}{m_a \vw} \FF_{ij}^s(\vw)  ,
}
with
\es{eq:F-c-s}{
\FF_{ij}^c(v) =  \int d^3 \vv f(\vv) \cos(m_a \vv \cdot \xx_{ij}) \delta[ |\vv|-v] , \\
\FF_{ij}^s(v) =  \int d^3 \vv f(\vv) \sin(m_a \vv \cdot \xx_{ij}) \delta[ |\vv|-v] .
}
By translation invariance, the entries of the correlation matrix only depend on the relative distances $\xx_{ij} \equiv \xx_i - \xx_j$.
These expressions then simplify for the correlations amongst a single detector, as $\FF_{ii}^c(v) = f(v)$ and $\FF_{ii}^s(v) = 0$.
But for $i \neq j$, the expressions in~\eqref{eq:F-c-s} contain a modulated version of the full velocity distribution, allowing us to extract non-trivial directional information about the velocity distribution $f(\vv)$ with multiple detectors separated by distances of order the de Broglie wavelength, where the integrand in~\eqref{eq:F-c-s} exhibits maximal variation. We note that the formalism we have developed assumes that the velocity distribution is stationary, or at least varies slowly on timescales compared to the axion coherence time. In Sec.~\ref{sec:daily} we will develop a formalism to take into account the daily modulation of $f(\mathbf{v})$ through a joint likelihood over multiple data-taking intervals.

%%%%%%%%%%%%%%%%%
\section{A Likelihood for Multi-Detector Axion Direct Detection}\label{sec:likelihood}
%%%%%%%%%%%%%%%%%

Having understood the statistics underlying the data collected by multiple detectors, we now outline how to incorporate these lessons into an appropriate likelihood.
The likelihood will be a simple generalization of the axion likelihood (generally applicable to wave-like DM) introduced in~\cite{Foster:2017hbq}, and we will closely follow their approach.
However, unlike in~\cite{Foster:2017hbq}, we work explicitly with the data as represented in $R_k$ and $I_k$ rather than the PSD, as the former notation exposes the full set of multi-detector correlations, as captured by $\bms$ in~\eqref{eq:fullcovariance}.
We will then outline how we can extract information about the parameters of $f(\vv)$ using this likelihood, exploiting where possible the asymptotic Asimov procedure~\cite{Cowan:2010js} to determine results analytically.
In Sec.~\ref{sec:parameters} we will then put the formalism to use in the context of several toy examples designed to highlight where interferometry opens up new avenues, and build intuition for the more realistic scenarios considered in Sec.~\ref{sec:daily}.

%%%%%%%%%%%%%%%%%
\subsection{The Multi-Detector Likelihood}
%%%%%%%%%%%%%%%%%

As detailed in Sec.~\ref{sec:stats}, we imagine we have a data set collected by $\mcn$ experiments, which each perform a time series of $N$ measurements collected at a frequency $f = 1/\Delta t$ of a quantity $\Phi \propto a$.
The real and imaginary part of the discrete Fourier transform of each experiments data set is constructed according to~\eqref{eq:re-im}, and then arranged into a single data set $d=\{\dd_0,\dd_1,\ldots,\dd_{N-1}\}$, with $\dd_k$ as given in~\eqref{eq:data_vector}.
We then define a model $\mathcal{M}$ with parameter vector $\bmt$ that has nuisance parameters $\bmt_{\rm nuis}$ describing the backgrounds in the individual experiments (encapsulated by $\lambda_{B,i}(\omega)$) and signal parameters $\bmt_{\rm sig}$ that characterize the axion contribution.
For example, $\bmt_{\rm sig}$ includes $\gagg$, $m_a$, and model parameters that describe the DM velocity distribution $f(\vv)$.
Then, as the data set is distributed according to a multivariate Gaussian, the appropriate likelihood is given by,
\es{eq:likelihood}{
\mathcal{L} (d | \mathcal{M}, \bmt) = \prod_{k=0}^{N-1} 
\frac{\exp \left[ - \frac{1}{2} \dd_k^T \cdot \bms_k^{-1}(\bmt) \cdot \dd_k  \right]}{\sqrt{(2 \pi)^{2 \mcn }|\bms_k(\bmt)|}} \,,
}
where $|\bms_k(\bmt)|$ is the determinant of the covariance matrix.

The utility of the likelihood function is that it facilitates inferences regarding the signal parameters, $\bmt_{\rm sig}$, from the data.
The ultimate goal of the axion DM program would be to infer a nonzero value of $A$, and hence the existence of a coupling between the Standard Model and DM, for example $\gagg$. 
Taking a frequentist approach to that problem, it is useful to define the following test statistic (TS) from the profile likelihood:
\es{eq:thetadef}{
\Theta(\bmt_{\rm sig}) = 2[ &\ln \mathcal{L}(d| \mathcal{M}, \{\hat{\bmt}_{\rm nuis}, \bmt_{\rm sig} \})  \\
- & \ln \mathcal{L}(d | \mathcal{M}, \{\hat{\bmt}_{\rm nuis}, \bmt_{\rm sig}={\bm 0}\})] .
}
In each likelihood, $\hat{\bmt}_{\rm nuis}$ denotes the value of the nuisance parameters that maximizes the likelihood for the given signal parameters.
The TS is then a function of the signal model parameters.
In particular, this means that in the first term in~\eqref{eq:thetadef} the nuisance parameters are uniquely determined at each $\bmt_{\rm sig}$ point by the values which maximize the log likelihood. 
The second term in~\eqref{eq:thetadef} is evaluated on the null model $\bmt_{\rm sig}={\bm 0}$, which can be achieved by setting the signal strength parameter $A$ to zero.

The TS in \eqref{eq:thetadef} is convenient for quantifying the significance of a putative signal, and we will use it throughout the following analysis.
In App.~\ref{app:stack} we describe a data-stacking procedure which reduces the data storage requirements for practical applications of our formalism.

%%%%%%%%%%%%%%%%%
\subsection{Asimov Test Statistic}
%%%%%%%%%%%%%%%%%

In order to build intuition for the information accessible to multiple detectors, we will use the Asimov data set~\cite{Cowan:2010js} to study the asymptotic TS analytically.
More precisely, the Asimov analogue of the TS in \eqref{eq:thetadef} is the average value taken over data realization,
\es{}{
\widetilde{\Theta}(\bmt_{\rm sig}) = \langle \Theta(\bmt_{\rm sig}) \rangle\,,
}
where the expectation value is taken on the data.
In order to evaluate the asymptotic TS, it is convenient to separate the model prediction, which enters through $\bms$, into background and signal contributions:
\es{}{
\bms = {\bf B} + {\bf S}\,.
}
Referring back to \eqref{eq:fullcovariance}, recall that the background is purely diagonal:
\es{}{
{\bf B} = \frac{1}{2}\, {\rm diag}(\lambda_{B,1},\,\lambda_{B,1},\,\ldots,\,\lambda_{B,\mcn},\,\lambda_{B,\mcn}).
}
Using this partitioning of the model prediction, we can then express the TS as follows
\es{eq:Theta-Asimov-step1}{
\Theta = \sum_{k=1}^{N-1} \left( \dd_k^T \left[ {\bf B}_k^{-1} - \bms_k^{-1} \right] \dd_k - \ln \left[ \frac{|\bms_k|}{|{\bf B}_k|} \right] \right).
}
Note that the values of ${\bf B}$ appearing in this expression are understood as being set to the value required by the profile likelihood technique.
In order to evaluate the Asimov form of this expression, we only need to evaluate the average on the first term, as the average is taken over the data.
This can be evaluated as follows,
\es{}{
&\left\langle \dd_k^T \left[ {\bf B}_k^{-1} - \bms_k^{-1} \right] \dd_k \right\rangle \\
= &{\rm Tr} \left( \left\langle \dd_k \dd_k^T \right\rangle \left[ {\bf B}_k^{-1} - \bms_k^{-1} \right]  \right),
}
and then as the data has mean zero, we know the above expected value is simply given by the true covariance matrix,
\es{}{
\left\langle \dd_k \dd_k^T \right\rangle
= \bms_k(\bmt=\bmt_{\rm truth}) = \bms_k^t\,.
}
Here the truth parameters can be considered as, for example, the parameters one would use when generating Monte Carlo to simulate expected experimental results.
For instance, to estimate the expected limit, the truth parameters would commonly have $A = 0$, whereas if we are estimating our sensitivity to features in $f(\vv)$, we will take $A \neq 0$ in the Asimov data.
In this work we are interested in the latter case, and therefore we will further assume the background has been fixed to the true value as a result of the profile likelihood technique,
\es{eq:AsimovTruth}{
\bms^t_k = {\bf S}_k^t + {\bf B}_k\,,
}
where ${\bf S}^t$ is the true signal model and the same ${\bf B}$ appears in both the Asimov and model predictions.
We further assume that the signal is always parametrically smaller than the background, which is the regime we will be in for any realistic experimental setup.\footnote{This assumption also ensures the validity of the fixed background being the same in {\it e.g.}~\eqref{eq:AsimovTruth} and~\eqref{eq:Theta-Asimov-step1}; if the signal is comparable to the background, then varying $A$ will generically alter the background determined by the profile likelihood technique.}
Implementing these assumptions, the Asimov form of \eqref{eq:Theta-Asimov-step1} is\footnote{To derive this result, the following identity is useful: $\ln |{\bf M}| = {\rm Tr} \ln {\bf M}$, for a matrix ${\bf M}$.}
\es{eq:Theta-Asimov-step2}{
\widetilde{\Theta}(\bmt_{\rm sig}) \approx \sum_{k=1}^{N-1} {\rm Tr} \left[ \left( {\bf S}_k^t - \frac{1}{2} {\bf S}_k \right) {\bf B}_k^{-1} {\bf S}_k {\bf B}_k^{-1} \right].
}

In \eqref{eq:Theta-Asimov-step2} we have a convenient form of the expected TS that is amenable to analytic study.
In the present work, our particular interest is the information contained in $f(\vv)$ that we can only access as a result of DM interferometry.
As such, it is convenient to evaluate a form of the Asimov TS, where all parameters except for those that control $f(\vv)$, as encoded in $\FF_{ij}^c(v)$ and $\FF_{ij}^s(v)$ in \eqref{eq:F-c-s}, are set to their true values in the presence of a non-zero signal.
If we further assume that our frequency resolution is sufficiently fine with respect to the scales over which the signal and background vary, then we can approximate the sum over Fourier components $k$ with an integral over frequencies $\omega$, or equivalently speeds $v = \sqrt{2 \omega/m_a-2}$.
Under these assumptions, the TS becomes
\begin{align}\label{eq:Asimov-fv}
\widetilde{\Theta} = &\frac{T\pi}{m_a} \int \frac{dv}{v}\, \sum_{i,j=1}^{\mcn} \frac{A_i A_j}{\lambda_{B,i} \lambda_{B,j}}
\left[ \FF_{ij}^c(v) \left( \FF_{ij}^{c,t}(v) - \frac{1}{2} \FF_{ij}^c(v) \right) \right. \nonumber\\
&\left.+ \FF_{ij}^s(v)\left( \FF_{ij}^{s,t}(v) - \frac{1}{2} \FF_{ij}^s(v) \right)
\right].
\end{align}
Much of the remainder of this work is devoted to studying the implications of this result.

%%%%%%%%%%%%%%%%%
\subsection{Limiting Cases of Zero and Infinite Separation}
%%%%%%%%%%%%%%%%%

We can use~\eqref{eq:Asimov-fv} to confirm basic asymptotic scalings expected for an analysis performed with DM interferometry.
To begin with, the Asimov TS for a single detector with response $A$ and background $\lambda_B$, recalling $\FF^s_{ii}(v)=0$ and $\FF^c_{ii}(v)=f(v)$, is given by
\es{eq:Theta-as-1detec}{
\widetilde{\Theta}_{\mcn = 1} = &\frac{A^2 T\pi}{m_a} \int \frac{dv}{v}\,\frac{f(v)}{\lambda_B^2} \left( f^t(v) - \frac{1}{2} f(v) \right).
}
This expression agrees with the result in~\cite{Foster:2017hbq}, which was derived for a single detector when analyzing the PSD.
Importantly, we emphasize once more that~\eqref{eq:Theta-as-1detec} is only dependent on the speed distribution, so directional parameters which affect the velocity distribution but not the DM speed distribution are inaccessible.
Before moving on to multiple detectors, we note that the expected discovery significance, which we denote TS, is given by the Asimov $\Theta$ evaluated at the model parameters that maximize the likelihood, which are the truth parameters.
Setting $f(v) = f^t(v)$ in~\eqref{eq:Theta-as-1detec} above gives
\es{eq:1-detec-discovery}{
{\rm TS}_{\mcn=1}  \approx \frac{A^2 T \pi}{2 m_a} \int \frac{dv}{v} \frac{f(v)^2}{\lambda_B^2}.
}

In order to extract directional parameters we need at least two detectors.
To that end, consider our expression in \eqref{eq:Asimov-fv} for $\mcn = 2$.
For simplicity, we take $A_1 = A_2 = A$ and $\lambda_{B,1} = \lambda_{B,2} = \lambda_B$, in which case the Asimov TS becomes
\es{eq:Theta-as-2detec}{
\widetilde{\Theta}_{\mcn=2} &= \frac{2 A^2 T \pi}{m_a} \int \frac{dv}{v\,\lambda_B^2} \left[  f(v) \left( f^t(v) - \frac{1}{2} f(v) \right) \right. \\
&+\FF_{12}^c(v) \left( \FF_{12}^{t,c}(v) - \frac{1}{2} \FF_{12}^c(v) \right)  \\
&+\left.\FF_{12}^s(v) \left( \FF_{12}^{t,s}(v) - \frac{1}{2} \FF_{12}^s(v) \right) \right].
}
In particular, the discovery TS is given by
\es{eq:TS-disc}{
{\rm TS}_{\mcn=2} = \frac{A^2 T \pi}{m_a} \int \frac{dv}{v} \frac{f(v)^2 + \FF_{12}^c(v)^2 + \FF_{12}^s(v)^2}{\lambda_B^2}.
}
Through $\FF^c_{12}$ and $\FF^s_{12}$, the discovery TS depends on the spatial separation of the two experiments $d = |\xx_{12}|$.
In the limit where the experiments are close with respect to the DM coherence length, {\it i.e.} $d \ll \lc$, then the two experiments see the same phase of the DM wave ($\kk \cdot \xx$ does not vary appreciably between them).
In this case, we would expect a coherent enhancement in the signal. Defining for future use
\es{eq:TS0}{
{\rm TS}_0 = \lim_{d \to 0} {\rm TS} = \frac{2A^2 T \pi}{m_a} \int \frac{dv}{v} \frac{f(v)^2}{\lambda_B^2},
}
we see from \eqref{eq:F-c-s} that for $\xx_{12}=0$ we have $\FF^c_{12}(v)=f(v)$ and $\FF^s_{12}(v)=0$, so  ${\rm TS}_0 = 4 {\rm TS}_{\mcn=1}\,.$
The $\mcn^2 = 4$ enhancement of the TS represents a coherent enhancement, a point emphasized in~\cite{Derevianko:2016vpm}.
This configuration provides a benchmark for the largest TS we can achieve for a general $\mcn=2$ configuration, and therefore will provide a convenient benchmark in the studies that follows.
On the other hand, for widely separated detectors with $d \gg \lc$, the DM fields will add incoherently.
For the problem at hand, again returning to \eqref{eq:F-c-s}, we see that the sine and cosine factors will oscillate rapidly, driving the integrals to zero.
What remains is,
\es{}{
\lim_{d \to \infty} {\rm TS} &= \frac{A^2 T \pi}{m_a} \int \frac{dv}{v} \frac{f(v)^2}{\lambda_B^2} = 2 {\rm TS}_{\mcn=1}\,,
}
so that the TS now only scales as $\mcn$, an incoherent enhancement.

The above argument can be readily generalized to $\mcn$ detectors.
Typically the signal strength $A$ is proportional to $\gagg^2$, so for $\mcn$ experiments all with pairwise separations $d \ll \lc$, we expect our sensitivity to $\gagg$ should scale coherently as $\mcn^{1/2}$.
If instead all experiments have $d \gg \lc$, the scaling is reduced to $\mcn^{1/4}$, and for scenarios outside these two extremes the scaling will be somewhere in between.
However, it is precisely this intermediate regime, where neither $\FF$ reduces to the speed distribution nor vanishes, where we expect to be able to extract additional information about $f(\vv)$.
We turn to the problem of estimating parameters of $f(\vv)$ in the context of the Asimov data set in the next section.

%%%%%%%%%%%%%%%%%
\section{Asimov Parameter Estimation}
\label{sec:parameters}
%%%%%%%%%%%%%%%%%

In this section we will use \eqref{eq:Theta-as-2detec} to perform frequentist parameter estimation and show explicitly that additional information about $f(\vv)$ can be extracted via DM interferometry.
For the purpose of simplifying the discussion, we will restrict our attention to the case of two detectors with equal background and detector responses as given in \eqref{eq:Theta-as-2detec}.
However, the entire discussion can be readily generalized to $\mcn$ arbitrary detectors by using the asymptotic TS expression in \eqref{eq:Asimov-fv}.

To be specific, imagine we are interested estimating a set of signal parameters $\bma$, which are a subset of the full set of signal parameters $\bma \subset \bmt_{\rm sig}$ related to $f(\vv)$, and which have true values $\bma^t$.
The Asimov procedure allows us to study our ability to infer these parameters.
For example, it is straightforward to confirm that $\widetilde{\Theta}(\bma)$ is maximized for $\bma = \bma^t$.\footnote{We emphasize that there is no guarantee that other parameters besides $\bma^t$ cannot also maximize the likelihood.
Indeed we will see exactly this possibility realized in a number of examples considered below.}
Beyond the best fit values, we are interested in determining the associated expected uncertainties and correlations between the various parameters, which are encompassed in the covariance matrix between the parameters, which we denote ${\bf C}$.
An estimator for this covariance matrix is given by the inverse Fisher information evaluated at the maximum likelihood, ${\bf C}^{-1} = {\bf I}(\hat{\bma})$ (again $\hat{\bma}$ are the parameters that maximize the likelihood), where
\es{eq:fisher}{
I_{ij}(\bma) = - \frac{\partial^2 \ln \mathcal{L}(\bma)}{\partial \alpha_i \partial \alpha_j} = - \frac{1}{2} \frac{\partial^2 \Theta(\bma)}{\partial \alpha_i \partial \alpha_j}\,,
}
where we use~\eqref{eq:thetadef}.
Given this relation, asymptotically our estimate for the covariance matrix is given by
\begin{align}
\label{eq:Cinv}
[\widetilde{\bf C}^{-1}]_{ij} 
= &- \frac{1}{2} \left.\frac{\partial^2 \widetilde{\Theta}(\bma)}{\partial \alpha_i \partial \alpha_j} \right|_{\bma = \bma^t} \\
= &\frac{A^2 T \pi}{m_a} \int \frac{dv}{v\,\lambda_B^2} \left[  
(\partial_i f(v)) (\partial_j f(v)) \right. \nonumber \\
+&\left.
(\partial_i \FF_{12}^{c}(v)) (\partial_j \FF_{12}^{c}(v))
+ (\partial_i \FF_{12}^{s}(v)) (\partial_j \FF_{12}^{s}(v)) \right]. \nonumber
\end{align}
This expression involves the following shorthand for derivatives of functions then evaluated at their truth values, $\partial_i = \partial/\partial \alpha_i |_{\alpha_i = \alpha_i^t}$.
The expression in the first line of this result lays bare a simple fact: if $\widetilde{\Theta}$ has no dependence on a particular parameter, for example the incident direction of a DM stream, or orientation of the Sun's motion through the DM halo, then the associated entries of the inverse covariance matrix vanish along with our ability to estimate that parameter.
For the case of a single parameter $\alpha$, we can readily invert the covariance matrix, and the above expression simplifies to
\es{eq:sig-a}{
\sigma_{\alpha}^{-2}
= &\frac{A^2 T \pi}{m_a} \int \frac{dv}{v\,\lambda_B^2} \left[  
(\partial_{\alpha} f(v))^2 \right.  \\
+&\left.
(\partial_{\alpha} \FF_{12}^{c}(v))^2
+ (\partial_{\alpha} \FF_{12}^{s}(v))^2 \right],
}
where again all parameters are evaluated at their truth values after derivatives.
We can already calibrate our basic expectation for parameter estimation from this result.
Optimal estimation of $\alpha$ amounts to maximizing the right hand side of the expression; indeed, as expected, increasing the signal strength, $A$, or the integration time, $T$, both achieve this.
If a parameter can be estimated from the speed distribution $f(v)$ (in other words, $\partial_{\alpha} f(v) \neq 0$), then that parameter may be estimated by a detector configuration with $d \ll \lc$.
However, the true power in the multi detector setup arises for parameters invisible to a single detector, defined by $\partial_{\alpha} f(v) = 0$, but where $\partial_{\alpha} \FF^{c,s}_{12}(v) \neq 0$.
In generic cases, such parameters are optimally estimated for $d \sim \lc$.

Continuing, let us assume that $\lambda_B$ is independent of frequency, in which case \eqref{eq:sig-a} becomes
\es{eq:sig-a-TS0}{
\sigma_{\alpha}^2
= &\frac{2}{{\rm TS}_0} \left[\int \frac{dv}{v} (f^t(v))^2 \right] \left\{\int \frac{dv}{v\,} \left[  
(\partial_{\alpha} f(v))^2 \right. \right. \\
+&\left.\vphantom{\frac{A^2 T \pi}{m_a} \int \frac{dv}{v\,\lambda_B^2}} \left.
(\partial_{\alpha} \FF_{12}^{c}(v))^2
+ (\partial_{\alpha} \FF_{12}^{s}(v))^2 \right] \right\}^{-1},
}
expressed in terms of ${\rm TS}_0$ as introduced in \eqref{eq:TS0}.
In particular, this result demonstrates the expected scaling of $\sigma_{\alpha} \sim ({\rm TS}_0)^{-1/2}$; the exact details will require a specific $f(\vv)$ and experimental configuration.
In the following subsections we will continue this line of thinking, demonstrating in several toy examples that a second detector can lift degeneracies from the single detector likelihood.

%%%%%%%%%%%%%%%%%
\subsection{The Minimal $\mcn=2$ Example}
%%%%%%%%%%%%%%%%%

\begin{figure}[t]
\centering	
\includegraphics[width=.47\textwidth]{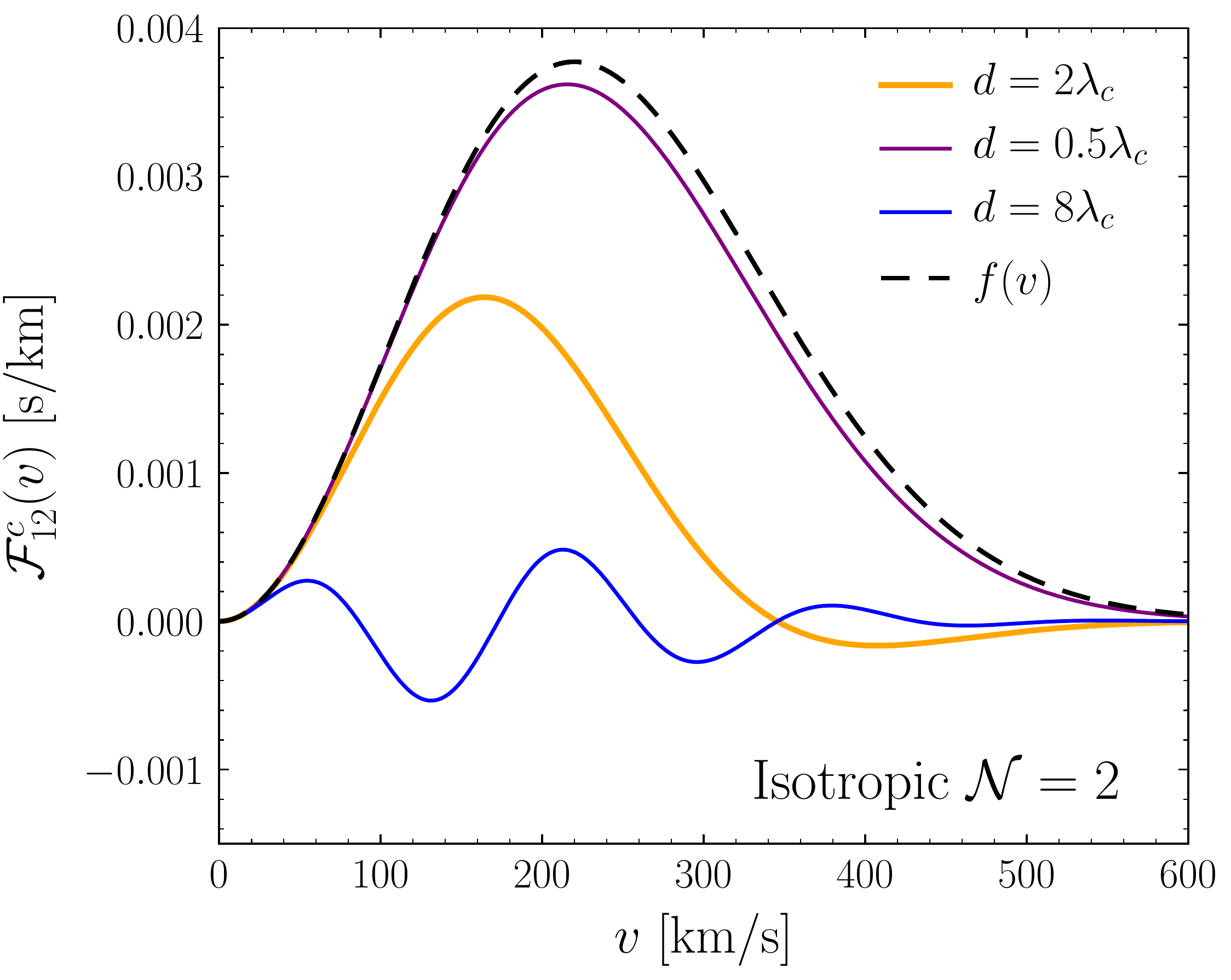}
\vspace{-0.3cm}
\caption{The modified speed distribution, $\FF^c_{12}(v)$, that carries the imprint of DM interferometry.
Here we show the particularly simple example of an isotropic SHM for $\mcn=2$ detectors, in which case the expression is given in \eqref{eq:F_simple_example}.
The result is shown for various choices of the two detector separation $d$ as compared to the axion coherence length $\lc = (m_a v_0)^{-1}$, with $v_0 = 220$ km/s.
The limiting cases of $\FF^c_{12}(v) \to f(v)$ for $d \ll \lc$ and $\FF^c_{12}(v) \to 0$ for $d \gg \lc$ are apparent.
For $d \sim \lc$, however, the profile is modulated with the interference inherent in the cross-spectrum.
In this simple case, there is no additional information about the velocity distribution that may be extracted by having multiple detectors.
}
\label{fig:isotropicNeq2}
\end{figure}

We begin our exploration of the above parameter estimation formalism with a simple scenario: $\mcn = 2$ detectors measuring DM drawn from an isotropic velocity in the laboratory frame, $4 \pi v^2 f({\bf v}) \equiv f(v)$.
This example is obviously idealized; in reality, the finite boost velocity of the Sun about the Galactic Center implies that even an isotropic velocity distribution in the Galactic frame will become anisotropic in the laboratory frame.
Nonetheless, this example will provide basic intuition for the impact of interferometry.

Invoking isotropy to perform the angular integrals, $\FF^{c,s}_{12}(v)$ can be computed as
\es{eq:F_simple_example}{
\FF_{12}^c(v)
&=f(v) \frac{\sin(m_a v d)}{m_a v d} \,, \qquad \FF_{12}^s(v) = 0 \,,
}
where again $d$ is the distance between the two detectors.
Thus for this example, we see explicitly that for $d \to 0$, we have $\FF_{12}^c(v) \to f(v)$, whereas for $d \to \infty$, instead $\FF_{12}^c(v) \to 0$.
As we will see in the examples below, it is the dispersion $v_0$ rather than the average speed $\bar{v}$ which determines the crossover between small and large $d$.

To progress further, we assume a concrete form for $f(v)$: the Maxwell-Boltzmann distribution,
\es{eq:MB}{
f(v) = {4 v^2 \over \sqrt{\pi} v_0^3} e^{-v^2 / v_0^2} \,,
}
where $v_0$ is the velocity dispersion.
Taking $v_0 \approx 220$ km/s, this velocity distribution is an approximation to the SHM that is expected to describe the bulk of the local DM, neglecting the finite velocity boost of the Sun relative to the Galactic Center, which breaks the isotropy in the laboratory frame. 
We will utilize the Maxwellian ansatz repeatedly in this work as an illustrative example.
In Fig.~\ref{fig:isotropicNeq2} we show $\FF_{12}^c(v)$ for various choices of $d/\lc$; there is a clear deviation from $f(v)$ when $d \sim \lc$, which is a manifestation of the nontrivial correlations in the multi-detector spectrum.
Note that we have defined, for the Maxwell-Boltzmann distribution, $\lc = (m_a v_0)^{-1}$, where $v_0$ is the velocity dispersion parameter that enters into~\eqref{eq:MB}; this is a particular realization of \eqref{eq:lcdef}.
Anticipating the more general scenario where the velocity distribution is not isotropic, it is precisely the deviation from $f(v)$ that we will use to extract information about the full velocity profile.

As we have chosen an isotropic $f(\vv)$, there is no additional information to extract about the velocity distribution in this case. Indeed, the distribution in \eqref{eq:MB} is defined by a single parameter, $v_0$, which we can envision estimating.
Evaluating \eqref{eq:sig-a-TS0} analytically in this case, we find
\es{}{
\frac{\sigma_{v_0}^2 {\rm TS}_0} {v_0^2}
= \frac{8\xi }{9\xi - \xi^3 + \sqrt{2}(15 + 2 \xi^2 + \xi^4) F[\xi/\sqrt{2}]}\,,
}
written in terms of a dimensionless distance scale $\xi = m_a v_0 d = d/\lc$, and Dawson's integral $F$.
We find explicitly that $\sigma_{v_0}$ is minimized for $\xi \to 0$, {\it i.e.} $d \ll  \lc$, since $\partial_{v_0} f(v) \neq 0$.

%%%%%%%%%%%%%%%%%
\subsection{The Infinitely-Cold Stream}
\label{sec:stream}
%%%%%%%%%%%%%%%%%

We now consider our first example of an anisotropic velocity distribution, a DM stream, and show that we can infer the direction of this stream using DM interferometry.
In addition to the bulk SHM, it is expected that the local DM velocity distribution could contain non-virialized substructure, such as cold tidal streams~\cite{Lee:2012pf,OHare:2014nxd,vanBibber:2003sv,Savage:2006qr,OHare:2017yze,Foster:2017hbq,Newberg:2003cu,Yanny:2003zu,Majewski:2003ux,Necib:2019zbk,OHare:2018trr,Myeong:2017skt,2018MNRAS.478.5449M}.
Streams are characterized by low velocity dispersions but large velocity boosts in the solar frame.
Let us suppose that in the laboratory frame the stream is boosted at velocity $\vv_{\rm str}$ and has velocity dispersion $v_0 \ll |\vv_{\rm str}|$.
In the limit $v_0 \to 0$, the velocity distribution approaches a delta function,
\es{eq:delta-stream}{
f(\vv) = \delta^3(\vv - \vv_{\rm str}) ,
}
which has an infinite coherence length but a finite de Broglie wavelength.
This is clearly an artificial example -- it is the maximally anisotropic velocity distribution -- but it is one we can evaluate fully analytically.
Further, a number of the conclusions that we will reach for the infinitely-cold stream will hold also in more realistic cases.
Note that for this example $f(v) = \delta(v-v_{\rm str})$, which has no dependence on the direction of the stream, and therefore a single detector cannot infer the direction.

As claimed, for this simple scenario, we can compute the exact global TS using~\eqref{eq:Theta-as-2detec}, and find
\es{eq:Theta-stream-example}{
\widetilde{\Theta}(\theta_{\rm str},\phi_{\rm str}) = {\rm TS}_0 \cos\left[m_a d v_{\rm str} (\hat \vv_{\rm str} - \hat \vv_{\rm str}^t) \cdot {\bf \hat x}_{12} \right].
}
We consider the TS as a function of the spherical coordinates of our test stream direction, $\bma = \{ \theta_{\rm str}, \phi_{\rm str}\}$, with the aim being to use the TS to infer the true direction of the stream, given by $\bma^t = \{ \theta_{\rm str}^t, \phi_{\rm str}^t\}$.
In this case we can also compute ${\rm TS}_0$, as defined in~\eqref{eq:TS0}, and we obtain\footnote{Note the fact that ${\rm TS}_0$ formally diverges,  $ {\rm TS}_0 \propto \delta(0)$, is an artifact of the stream having a delta-function speed distribution.
The divergence is regulated by the finite dispersion of the stream, as we discuss below.}
\es{}{
{\rm TS}_0 = \frac{2A^2 T \pi}{m_a \lambda_B^2} \frac{\delta(v_{\rm str}-v_{\rm str})}{v_{\rm str}}\,.
}
Now consider the angle-dependent factor in \eqref{eq:Theta-stream-example}.
Without loss of generality, we take $\hat{\xx}_{12} = \hat{\bf z}$ and define spherical coordinates with respect to $\hat{\xx}_{12}$, so that the argument of the cosine in \eqref{eq:Theta-stream-example} simplifies to
\es{eq:vdotx-str}{
(\hat{\vv}_{\rm str}-\hat{\vv}_{\rm str}^t) \cdot \hat{\xx}_{12} 
= \cos \theta_{\rm str} - \cos \theta_{\rm str}^t\,,
}
where $\theta_{\rm str}$ and $\theta_{\rm str}^t$ are the usual polar angles in spherical coordinates.
Neither azimuthal coordinate $\phi$ appears in this expression, and hence the azimuthal angles are also absent in the TS.
This implies we cannot infer one of the angular coordinates of $\vv_{\rm str}^t$ from the data.
For our particular choice of coordinates, we can infer the parameter $\theta_{\rm str}^t$, as we will describe below, but the likelihood has a flat direction in $\phi_{\rm str}$, so that we cannot infer the associated truth value.
The degeneracy is physical.
In our coordinates the symmetry of the likelihood is represented by an invariance under changes in $\phi$, but more generally the TS is unchanged by rotations about the detector separation axis, $\hat{\xx}_{12}$.
This can be seen from the dependence of the TS on $(\hat{\vv}_{\rm str}-\hat{\vv}_{\rm str}^t) \cdot \hat{\xx}_{12}$: any change in the test or true $\hat{\vv}_{\rm str}$ that is perpendicular to $\hat{\xx}_{12}$ has no impact.

This symmetry of the TS under rotations around $\hat{\xx}_{12}$ is in fact not a relic of our idealized example.
Our ability to infer the direction of a velocity parameter vector that defines a given $f(\vv)$ enters through the $\vv \cdot \xx_{12}$ in $\FF^{c,s}_{12}$.
But as $\vv \cdot \xx_{12}$ is itself invariant to rotations of the velocity about the $\hat{\xx}_{12}$ axis, one can show that this flat direction in the likelihood exists generally -- indeed we will see it in more realistic cases (a direct analogue is apparent in the symmetry observed in Fig.~\ref{fig:toymap}, related to the SHM example discussed below).
This symmetry will be broken by a dependence in the likelihood on multiple detector axes that are not parallel, provided either by a third detector or alternatively by daily modulation, where the single $\hat{\xx}_{12}(t)$ will vary throughout the day at different times $t$.
We will explore this latter example in detail in Sec.~\ref{sec:daily} -- indeed the optimal detector configuration will be determined by maximally violating this symmetry -- but until then the symmetry will represent a basic feature of the physics.

Returning to our specific coordinate system where $\hat{\xx}_{12} = \hat{\bf z}$, we may perform parameter estimation on the angle between the stream and detector.
From \eqref{eq:sig-a-TS0}, we have
\es{eq:delta_function}{
\sigma_{\theta_{\rm str}}^{2} &= \frac{2}{{\rm TS}_0} \frac{1}{(m_a v_{\rm str} d)^2} \frac{1}{\sin^2 \theta^t_{\rm str}}\,.
}
Note that the uncertainty on the parameter $\theta_{\rm str}$ is minimized for $\theta^t_{\rm str} = \pi/2$, { \it i.e.} when the stream is perpendicular to the two-detector axis.
On the other hand, if the two vectors are parallel, $\theta^t_{\rm str} = 0$ or $\pi$, then we see $\sigma_{\theta_{\rm str}}$ diverges.
Yet we can still infer $\theta_{\rm str}^t$ in this case.
Indeed, looking to \eqref{eq:vdotx-str} we see that the asymptotic TS depends on $\theta_{\rm str}$; the likelihood is not globally flat, and we can estimate the angle from contours around the maximum likelihood.
Instead, in this case there is a breakdown of the quadratic approximation around the maximum likelihood.
If we were to incorporate higher derivatives than in \eqref{eq:Cinv}, we would confirm that the likelihood is not truly flat at these points.
This of course should be contrasted with the true flat direction in the likelihood associated with $\phi_{\rm str}$.
Note, however, that as $\theta_{\rm str}^t$ approaches either $0$ or $\pi$, becoming parallel to $\xx_{12}$, the undetermined parameter $\phi_{\rm str}$ is less relevant.
In the limit where the two vectors are parallel, we can infer the true direction of the stream, in spite of this degeneracy.

There is another interesting feature in~\eqref{eq:delta_function}: the result suggests that we can take $d \to \infty$ to constrain this one direction of the stream to arbitrary precision.
This is a manifestation of our assumption that the stream has no velocity dispersion: it remains coherent over arbitrary large distances, allowing for an improved baseline over which we can measure the stream direction.
To study this feature further, imagine making this example slightly more realistic by introducing a finite velocity dispersion $v_0$, with $v_0 \ll v_{\rm str}$, such that $f(\vv)$ has support in a small volume of radius $\sim v_0$ around $\vv_{\rm str}$.
For small enough $v_0$ we would expect the results of the $\delta$-function stream to hold.
Yet there is an important conceptual difference: the coherence length is no longer infinite because the different waves that constitute the local DM field now have speeds that vary by $\mathcal{O}(v_0)$.
Parametrically, the argument of the interferometric terms scale as $m_a |\vv| |\xx_{12}| \sim m_a d (v_{\rm str} + \mathcal{O}(v_0))$, but with the $\mathcal{O}(v_0)$ term varying between states.
If we now take $d \gg (m_a v_0)^{-1}$, then the different waves will add incoherently, suppressing the power.
But if we choose $d \sim (m_a v_0)^{-1}$, a degree of coherence can be maintained, along with the interference pattern carrying the information we seek to extract (see also the orange curve $d = 2\lc$ in  Fig.~\ref{fig:isotropicNeq2}).
Accordingly, for the optimal separation, the scaling of the sensitivities in \eqref{eq:delta_function} is (taking $\sin^2 \theta_{\rm str}^t \sim 1/2$ for definiteness)
\es{eq:TSv0scale}{
\sigma_{\theta} \sim \frac{2}{\sqrt{{\rm TS}_0}} \frac{v_0}{v_{\rm str}} = \frac{2}{\sqrt{{\rm TS}_0}} \frac{\ldB}{\lc}\,.
}

In the more realistic examples we will confirm the conclusion that $d \sim \lc$ provides the maximum sensitivity.
There is another consequence of this choice.
Taking $d = (m_a v_0)^{-1}$ in \eqref{eq:Theta-stream-example}, the prefactor of the dot product in \eqref{eq:vdotx-str} is $v_{\rm str}/v_0 \gg 1$, by definition of this being a cold stream.
Small variations in $(\hat{\vv}_{\rm str}-\hat{\vv}_{\rm str}^t) \cdot \hat{\xx}_{12}$ will induce large variations in the argument of the cosine, implying that the global structure of the TS is highly nontrivial.
Although the maximum TS will be attained at the true $\theta$, there will be a pattern of local maxima with comparable TS (this result is depicted in Fig.~\ref{fig:sgr_example}, and persists even with daily modulation as shown in Fig.~\ref{fig:SGR_Modulation}).

%%%%%%%%%%%%%%%%%
\subsection{The (boosted) Standard Halo Model}
\label{sec:SHM}
%%%%%%%%%%%%%%%%%

The bulk DM halo of the Milky Way is expected to be Maxwell-Boltzmann distributed as in~\eqref{eq:MB} in the Galactic frame, except for a possible cut-off around the escape velocity $\sim$500 km/s~\cite{Piffl:2013mla}. 
On the other hand, the Sun is boosted with respect to the Galactic frame by~\cite{2010MNRAS.403.1829S} 
\es{}{
\vv_{\odot} \approx (11,\,232,\,7)~{\rm km/s} \,,
}
in Galactic coordinates, where ${\bf \hat x}$ points towards the Galactic Center, ${\bf \hat y}$ points in the direction of the local rotation of the disk, and ${\bf \hat z}$ points towards the Galactic north pole.
Thus in the laboratory frame (neglecting the Earth's motion), the velocity distribution becomes that of the SHM,
\es{eq:SHM_full}{
f({\bf v}) = \frac{1}{\pi^{3/2} v_0^3} e^{-(\vv + \vv_\odot)^2 / v_0^2} \,,
}
with a velocity dispersion $v_0 \approx 220$ km/s~\cite{Freese:2012xd,Herzog-Arbeitman:2017fte}.
Note in particular that $v_0 \sim |\vv_\odot| \equiv v_\odot$, so for the SHM $\lc \sim \ldB$.
The associated speed distribution is
\es{eq:SHM}{
f(v) = \frac{v}{\sqrt{\pi} v_0 v_{\odot}} e^{-(v+v_\odot)^2/v_{0}^2} \left( e^{4 v v_{\odot}/v_0^2} - 1 \right).
}
As we have emphasized many times already, single detectors are only sensitive to the speed distribution, which only depends on $v_\odot$ but not the orientation of the solar velocity $\hat \vv_\odot$.
Thus, a single detector may constrain the model parameters $v_0$ and $v_\odot$ (as shown in~\cite{Foster:2017hbq}), but determining the orientation requires multiple detectors.\footnote{In principle annual modulation may be used by a single detector to infer $\hat \vv_\odot$, as discussed in~\cite{OHare:2017yze,Foster:2017hbq}.}

To determine the expected sensitivity to the direction $\hat \vv_\odot$ we need to compute the derivatives of $\FF^{c,s}_{12}(v)$ that appear in \eqref{eq:sig-a-TS0}:
\begin{widetext}
\es{eq:SHM_details}{
\left.\partial_{\theta_{\odot}}\FF_{12}^c(v) \right|_{\theta_{\odot} = \theta_{\odot}^t}
=&\,\frac{4v^3 v_{\odot} e^{-(v^2+v_\odot^2)/v_0^2}}{\sqrt{\pi} v_0^5} \int d\theta \sin \theta\, \exp \left[ -\frac{2 v v_{\odot}}{v_0^2} \cos \theta \cos \theta_{\odot}^t \right]
\cos(m_a v d \cos \theta) \\
\times& \left[ I_0 \left[ \frac{2 v v_{\odot}}{v_0^2} \sin \theta \sin \theta_{\odot}^t \right] \cos \theta \sin \theta_{\odot}^t + I_1 \left[ \frac{2 v v_{\odot}}{v_0^2} \sin \theta \sin \theta_{\odot}^t \right] \sin \theta \cos \theta_{\odot}^t \right],
}
\end{widetext}
where $I_{0,1}$ are both modified Bessel functions (an analogous expression holds for $\FF^{s}_{12}$).
In computing this result we have again chosen coordinates $\hat{\xx}_{12} = \hat{\bf z}$, but left the direction of $\hat{\vv}_{\odot}$ arbitrary, defined by $(\theta_{\odot}^t,\phi_{\odot}^t)$.
The most important feature of this result is that it exhibits no dependence upon $\phi_{\odot}^t$: again, there is a symmetry in the likelihood for rotations around $\hat{\xx}_{12}$.
Beyond this, we can also see that the derivative vanishes when $\vv_{\odot}$ is parallel to the detector separation ($\theta_{\odot}^t=0$).
Accordingly, in this case we will find $\sigma_{\theta_{\odot}}$ diverges, as we did for the stream.
But again this is not a global flat direction in this case; the likelihood is just sufficiently flat at the maximum that the first three derivatives vanish.

\begin{figure}[!t]
\centering	
\includegraphics[width=.47\textwidth]{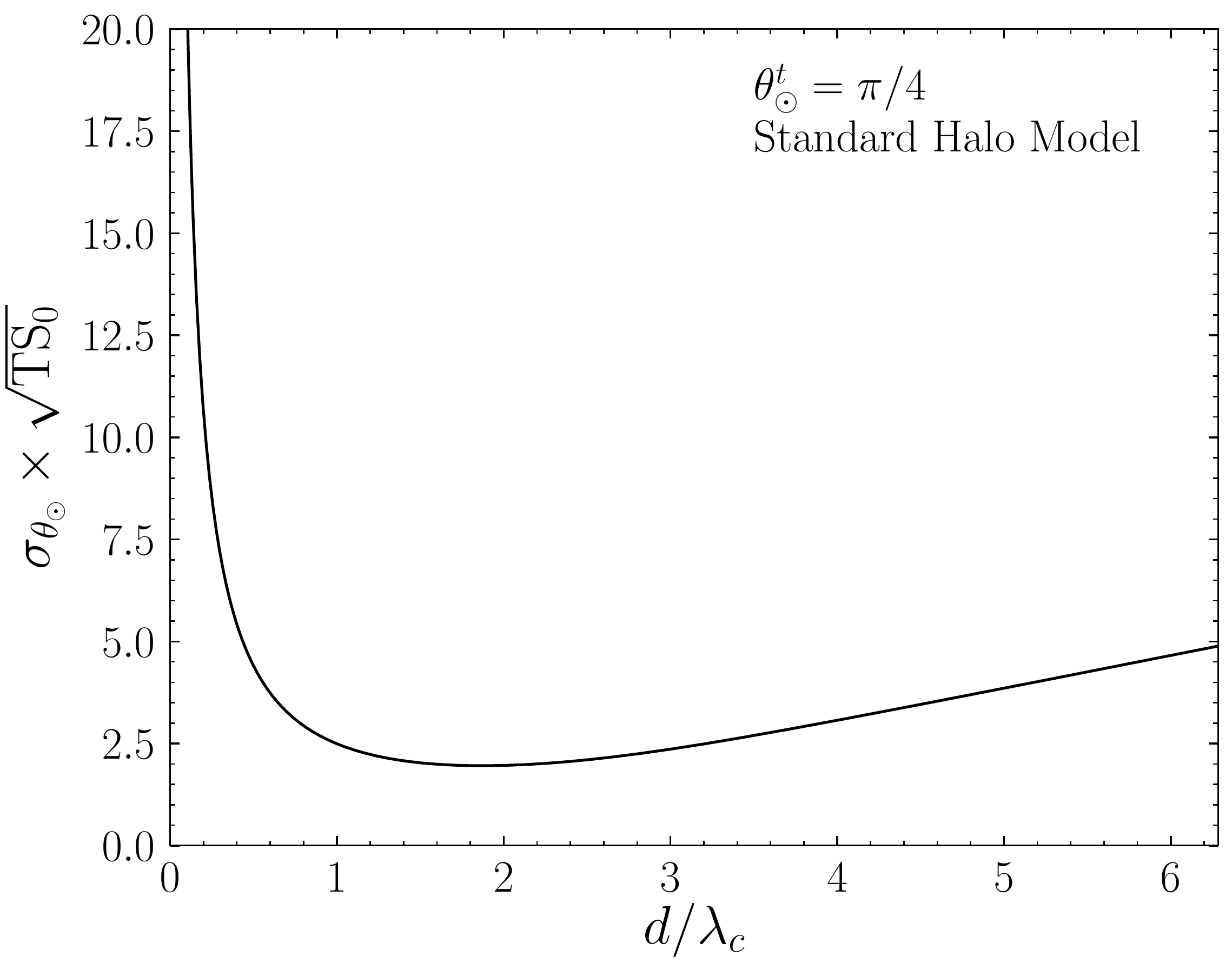}
\vspace{-0.3cm}
\caption{The expected uncertainty on the angle between the detector axis and solar velocity, $\theta_\odot = \arccos(\hat{\vv}_{\odot} \cdot \hat{\xx}_{12})$, as a function of $d/\lc = d \times m_a v_0$.
In this example we have set the true orientation to $\theta_\odot^t = \pi/4$.
With this configuration, we find that the maximum precision is obtained for $d  \approx 2 \lc$.
}
\label{fig:SHM_example}
\end{figure}

\begin{figure*}[htb]
\centering
\includegraphics[width=.44\textwidth]{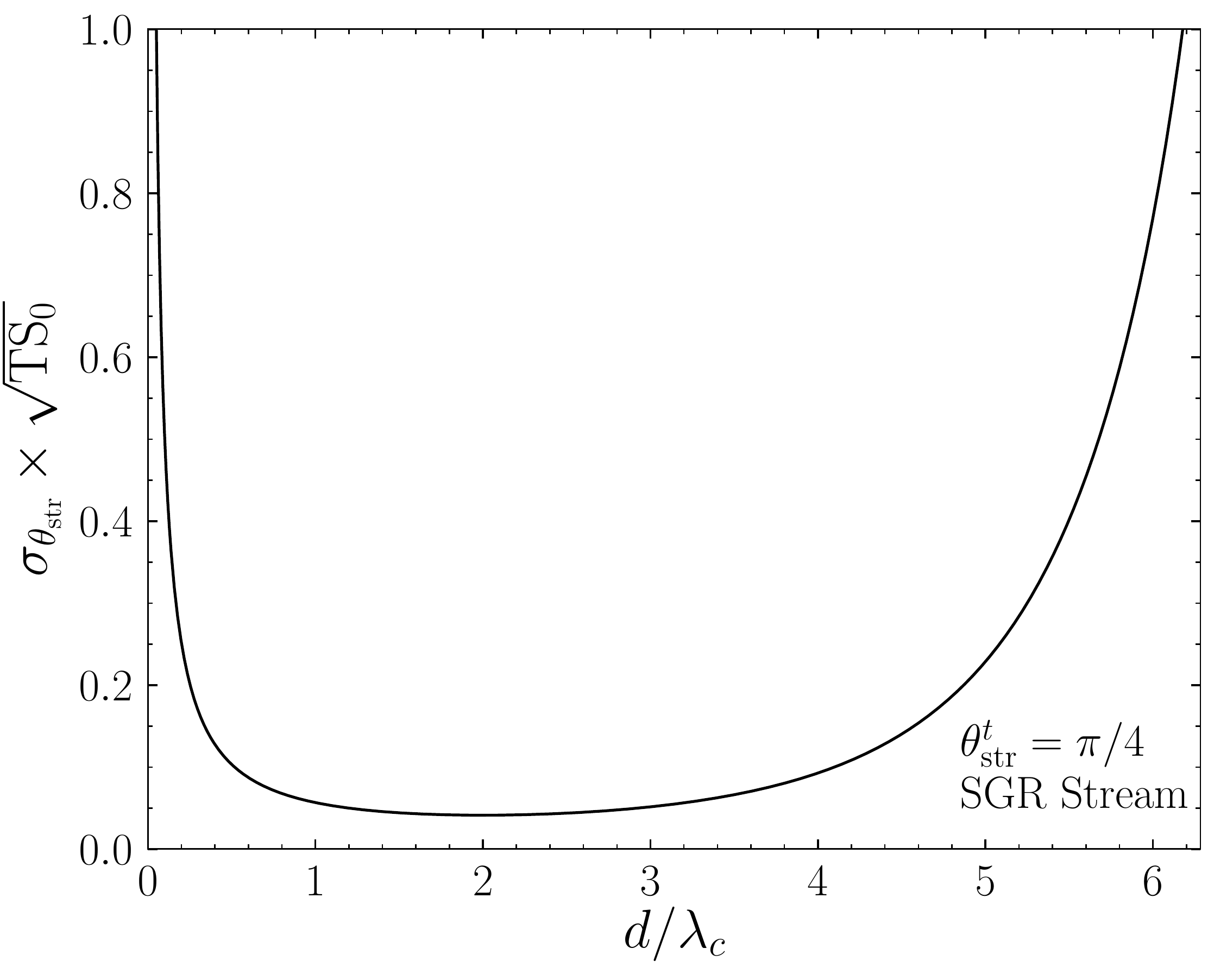}
\hspace{0.5cm}
\includegraphics[width=.47\textwidth]{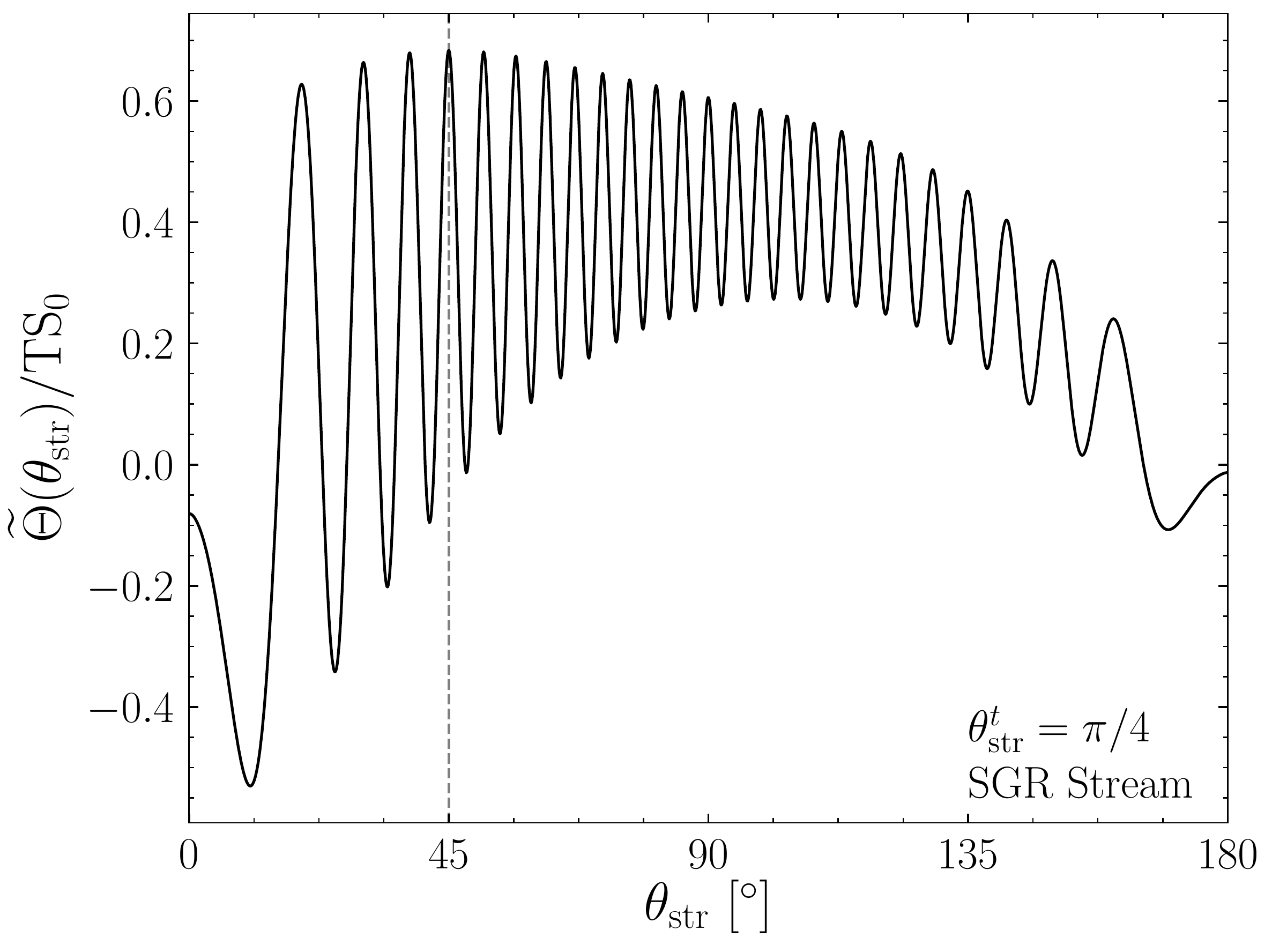}
\vspace{-0.3cm}
\caption{(\textit{Left}) As in Fig.~\ref{fig:SHM_example}, but for the Sagittarius (SGR) stream rather than for the SHM.
As before, the maximum precision for the inferred value of $\theta_{\rm str}$ is achieved at $m_a v_0 d \approx 2$, although the overall dependence is somewhat softened outside of the extremes at $m_a v_0 d = 0$ and $m_a v_0 d = 2 \pi$.
The values of $\sigma_{\theta_{\rm str}} \times \sqrt{\mathrm{TS}_0}$ are also considerably smaller than those found in the SHM example, indicating that the angle $\theta_{\rm str}$ can be reconstructed with much greater precision for the SGR stream as compared to the SHM.
(\textit{Right}) The Asimov TS $\widetilde \Theta(\theta_{\rm str})$ for the SGR stream rescaled by the co-located detection significance $\mathrm{TS}_0$ as a function of $\theta_{\rm str}$ for a detector configuration where the true stream direction is $\theta_{\rm str}^t = \pi/4$ (dashed vertical line).
We have fixed $m_a v_0 d = 2$.
The TS $\widetilde \Theta(\theta_{\rm str})$ is maximized at the true value of $\theta_{\rm str}$, but there is considerable nontrivial global structure with a large number of local minima and maxima in $\widetilde \Theta$.} 
\label{fig:sgr_example}
\end{figure*}

To proceed beyond these analytic insights, we will compute the remaining results numerically.
We define the angle between $\vv_{\odot}$ and $\xx_{12}$ as $\theta_\odot$.
To begin with, we take a generic value of $\theta_\odot^t = \pi/4$ and consider how well we can infer this angle as a function of detector separation.
The results are shown in Fig.~\ref{fig:SHM_example}.
Unlike for the $\delta$-function stream, there is now a minimum at a finite value of $d$, and as argued on general grounds this occurs when $d \sim \lc = (m_a v_0)^{-1}$.
That the uncertainty diverges for $d \to 0$ is consistent with the fact that a single detector cannot infer this direction.
In more detail we find the minimum occurs at $d \sim 2 \lc$, where we obtain $\sigma_{\theta_d} \approx 2/\sqrt{{\rm TS}_0}$.
For example, if ${\rm TS}_0 = 25$, corresponding to a 5$\sigma$ local significance detection with $d = 0$, then at the distance $d \sim 2\lc$ corresponding to minimum uncertainty, the solar velocity direction with respect to the detector axis could be localized to $0.4~{\rm rad} \sim 20^\circ$ on the sky.
We can understand the magnitude of $\sigma_{\theta_d}$ at its minimum from~\eqref{eq:TSv0scale}: the SHM has the form of a stream where $v_0 \sim v_{\rm str} = v_{\odot}$, and therefore we would expect $\sigma_{\theta_d} \times \sqrt{{\rm TS}_0} \sim 2$, exactly as observed.

\begin{figure*}[htb]
\centering	
\includegraphics[width=.99\textwidth]{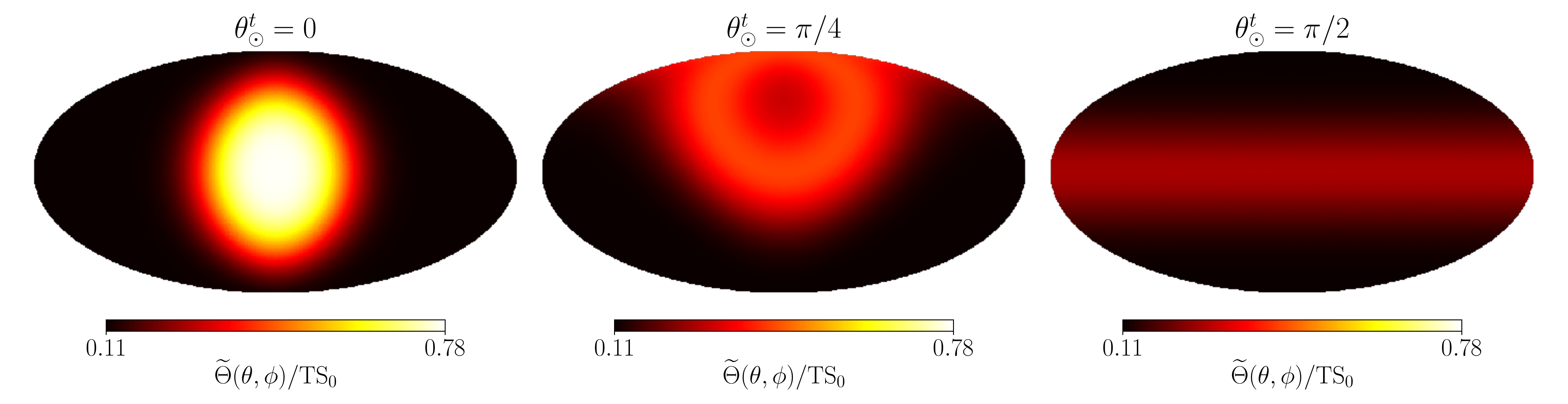} 
\vspace{-0.3cm}
\caption{(\textit{Left}) A Mollweide projection of the Asimov test statistic $\widetilde \Theta(\theta, \phi)$ for the SHM divided by the co-located detection significance $\mathrm{TS}_0$.
The detectors are configured so that the displacement vector between them is parallel to the SHM boost velocity, and the Mollweide plot is rotated so that it is centered on the maximum test statistic.
(\textit{Center}) As on the left, but for a detector configuration where the displacement vector is at a $45^\circ$ angle to the North ($\theta_\odot^t = \pi/4$) with respect to the SHM boost velocity.
(\textit{Right}) As on the left, but for a detector configuration where the displacement vector is perpendicular to the SHM boost velocity ($\theta_\odot^t = \pi/2$).
In this configuration the location of the boost velocity can only be localized to a great circle on the celestial sphere.}
\label{fig:toymap}
\end{figure*}

However, as we have emphasized already, it is important to keep in mind that our estimate of $\sigma_{\theta_\odot}$ is a measure of the expected curvature of the likelihood in the vicinity of the true value and does not capture the global structure of the expected likelihood function.
To illustrate these features we fix $d = 2 \lc$ for definiteness and illustrate the global map $\widetilde{\Theta}(\theta_\odot,\phi_\odot)/{\rm TS}_0$ in Fig.~\ref{fig:toymap}, for three different values of $\theta_\odot^t$.
Note that we have divided out the overall significance ${\rm TS}_0$, so that exactly how well we can localize the direction will depend on how significantly the DM signal has been measured.
However, the expected global structure of the TS will be a rescaled version of these maps.
In each case the true $\hat{\vv}_{\odot}$ that we are seeking to infer is located in the center of the Mollweide projection maps.
The left panel illustrates the scenario with $\hat{\xx}_{12} = \hat{\vv}_{\odot}$ ($\theta_\odot=0$), the center has $\theta_\odot = \pi/4$, while in the right panel two directions are perpendicular and $\hat{\xx}_{12}$ points between the poles of the map ($\theta_\odot=\pi/2$).
In all cases the symmetry of the TS around the $\hat{\xx}_{12}$ axis is apparent.
The only case where this flat direction in the maximum TS is not an obstruction to determining the true direction of $\hat{\vv}_{\odot}$ is when $\theta_d=0$.
In that case we are still able to localize the true direction, although we note the likelihood is relatively flat around the maximum (consistent with the second derivative vanishing).
In Sec.~\ref{sec:daily} we illustrate how daily modulation generically allow us to fully determine both of the angles associated with the direction of $\vv_\odot$.

%%%%%%%%%%%%%%%%%
\subsection{The Sagittarius Stream}
%%%%%%%%%%%%%%%%%

As a final example working with a single static $\hat{\xx}_{12}$, we return to the case of the cold stream with non-vanishing velocity dispersion.
We expect many of the conclusions reached in Sec.~\ref{sec:SHM} to hold in this case. In particular, the symmetry around the $\hat{\xx}_{12}$ axis will remain, but we will see explicitly in this case the non-trivial structure induced in the global likelihood by the ratio of $v_0/v_{\rm str} \sim \ldB/\lc \gg 1$.
To make the example concrete, the DM component of the Sagittarius stream may extend to the Sun's location, and estimates~\cite{Vogelsberger:2008qb,2011MNRAS.415.2475M} suggest that it could make up $\sim$5\% of the local DM density.
However, the DM associated with the stream would be highly collimated in phase space; we follow~\cite{Foster:2017hbq} and model the Sagittarius stream DM velocity distribution by a boosted Maxwellian as in \eqref{eq:SHM_full}, but with  $v_0 = 10$ km/s and $\vv_{\odot}$ replaced by $\vv_{\rm str} = (0, 93.2, -388)$ km/s~\cite{Savage:2006qr,OHare:2014nxd}.
We consider the stream in isolation, as opposed to in conjunction with the bulk SHM DM phase-space distribution, because even though the stream component is sub-dominant in terms of DM density, it still dominates in the narrow region of phase space where the Sagittarius stream has compact support.
To simplify the discussion we will simply take $v_{\rm str} = 400$ km/s, with a direction that we will again specify by its angle with respect to the detector axis (given the degeneracy in rotations about that axis).
Note that this example could apply equally well to other putative DM streams, such as the newly discovered S1 stream~\cite{Myeong:2017skt,2018MNRAS.478.5449M,OHare:2018trr}.

To begin with, in the left panel of Fig.~\ref{fig:sgr_example} we show the expected uncertainty on the recovered angle between the stream and detector, $\theta_{\rm str}$, as a function of the distance in units of $\lc = (m_a v_0)^{-1}$, for a true value $\theta_{\rm str}^t = \pi/4$.
This figure is the stream analogue of what we showed for the SHM in Fig.~\ref{fig:SHM_example}.
Once more, following the general discussion in Sec.~\ref{sec:stream}, the optimal sensitivity is achieved for $d \sim \lc$, and from \eqref{eq:TSv0scale}, we expect $\sigma_{\theta_d} \times \sqrt{{\rm TS}_0} \sim 2 v_0 / v_{\rm str} \sim 0.05$ at the minimum-uncertainty distance, compatible with what we see in Fig.~\ref{fig:sgr_example}.

However, just like in the case of the SHM it is important to also examine the global properties of the TS in addition to the curvature of the expected TS at the true parameter values.
Towards that end, on the right of Fig.~\ref{fig:sgr_example} we show the expected TS $\widetilde{\Theta}$, normalized to ${\rm TS}_0$, as a function of the reconstructed angle between the stream and detector, $\theta_{\rm str}$.
For this figure we have fixed the true orientation at $\theta_{\rm str}^t = \pi / 4$ along with the separation $d = 2 \lc$.
We see that $\widetilde{\Theta}$ drops off quickly around the true value of $\theta_{\rm str} = \pi/4$ (vertical dashed), but that there is non-trivial structure with local maxima at larger and smaller $\theta_{\rm str}$ values.
This is a direct manifestation of the non-trivial interference patterns discussed in Sec.~\ref{sec:stream} for cold streams: the large ratio $v_{\rm str}/v_0$ enters into the argument of the trigonometric functions in $\FF^{c,s}_{12}(v)$.

%%%%%%%%%%%%%%%%%
\section{Daily modulation}\label{sec:daily}
%%%%%%%%%%%%%%%%%

One of the most dramatic signatures of DM interferometry is the unique daily modulation signal available to multiple detectors.
This effect, which we describe in the current section, would be a smoking gun signature that an emerging excess has a DM origin, and it also allows two detectors to better determine geometric parameters describing the velocity distribution.
The basic idea is simply that for two detectors fixed at generic locations on the surface of the Earth, the separation vector $\xx_{12}$ is rotating in the inertial Galactic frame throughout the day.
This is in contrast to the angular parameters entering in the DM velocity distribution, such as the Solar direction $\hat \vv_\odot$, which should always point in the same Galactic direction, regardless of the orientations of the detectors at any point in time on Earth.
The rotation of $\xx_{12}$ with respect to the fixed $\hat \vv_\odot$ implies that we will sample a variety of angles between the two vectors, and therefore vary the modulation of the speed distributions in $\FF^{c,s}_{12}(v)$, as already depicted in Fig.~\ref{fig:exampleF}.
Critically this will lift the flat direction in the maximum likelihood associated with rotations around $\xx_{12}$ that we observed repeatedly in Sec.~\ref{sec:parameters}: as the likelihood will now depend on a collection of different vectors $\xx_{12}(t)$ , the symmetry that exists around any one of them will not be preserved in the full TS.

In the rest of this section we divide the discussion into three parts.
Firstly, we describe how to construct the likelihood for the generic case of $\mcn$ detectors incorporating daily modulation and describe how it is straightforward to generalize our full formalism to this case.
We then focus on the specific case of $\mcn = 2$ and show, within the Asimov formalism, how the examples of the SHM and Sagittarius stream discussed above are modified in the presence of daily modulation.
Finally, we turn to a Monte Carlo simulation of a realistic example and demonstrate how, within a day, a resonant experiment could constrain the direction of the solar velocity vector, $\vv_{\odot}$, that controls the SHM to sub-degree accuracy.

%%%%%%%%%%%%%%%%%
\subsection{A Likelihood with Daily Modulation}
%%%%%%%%%%%%%%%%%

So far in this work, we have envisioned a set of $\mcn$ experiments collecting measurements of the signal-plus-background frequency spectra for a duration of time $T$ while the detector separations were fixed with respect to the boost velocity of the DM component under consideration.
However, this framework cannot be extended to the case of daily modulation, as the signal prediction will fundamentally be varying over a 24-hour period.
In order to properly account for this effect, the data must be collected in time intervals of duration $T \ll 24$ hours and analyzed with a joint likelihood over all the collected intervals.
In detail, if we imagine that we collect $M$ such time intervals, indexed by $r=0,1,\ldots,M-1$, then for each of these we will have a data set $d_r = \{\dd_{k,r}\}$,  where again $k$ labels the Fourier mode.
For each data set $d_r$, we can compute the likelihood as in \eqref{eq:likelihood}, and the full joint likelihood is the product of these over $r$.
Explicitly, we have
\begin{align}
\label{eq:likelihood-daily}
\mathcal{L} (d | \mathcal{M}, \bmt) 
= &\prod_{r=0}^{M-1} \mathcal{L} (d_r | \mathcal{M}, \bmt) \\
= &\prod_{r=0}^{M-1}\prod_{k=0}^{N-1} 
\frac{\exp \left[ - \frac{1}{2} \dd_{k, r}^T \cdot \bms_{k, r}^{-1}(\bmt) \cdot \dd_{k, r}  \right]}{\sqrt{(2 \pi)^{2 \mcn }|\bms_{k, r}(\bmt)|}}. \nonumber
\end{align}
Importantly, note that we have also attached an index $r$ to the signal prediction $\bms(\bmt)$, as we need to account for the variation of the detector separations $\xx_{ij}$ throughout the day.

In a similar fashion the full formalism of Secs.~\ref{sec:likelihood} and~\ref{sec:parameters} can be extended to include the varied detector orientation: within a given sub-interval we simply adjust $\xx_{ij}$ as appropriate, and then we form joint quantities by combining these as in the likelihood above.
To provide just a single illustrative example, the Fisher information computed in \eqref{eq:fisher}, would become
\es{}{
I_{ij}(\bma) =  - \frac{1}{2}\sum_{r=0}^{M-1} \frac{\partial^2 \Theta_r(\bma)}{\partial \alpha_i \partial \alpha_j}\,,
}
with other expressions similarly generalized.

%%%%%%%%%%%%%%%%%
\subsection{Asimov Examples with Daily Modulation}
%%%%%%%%%%%%%%%%%

\begin{figure*}[htb]
\centering	
\includegraphics[width=.99\textwidth]{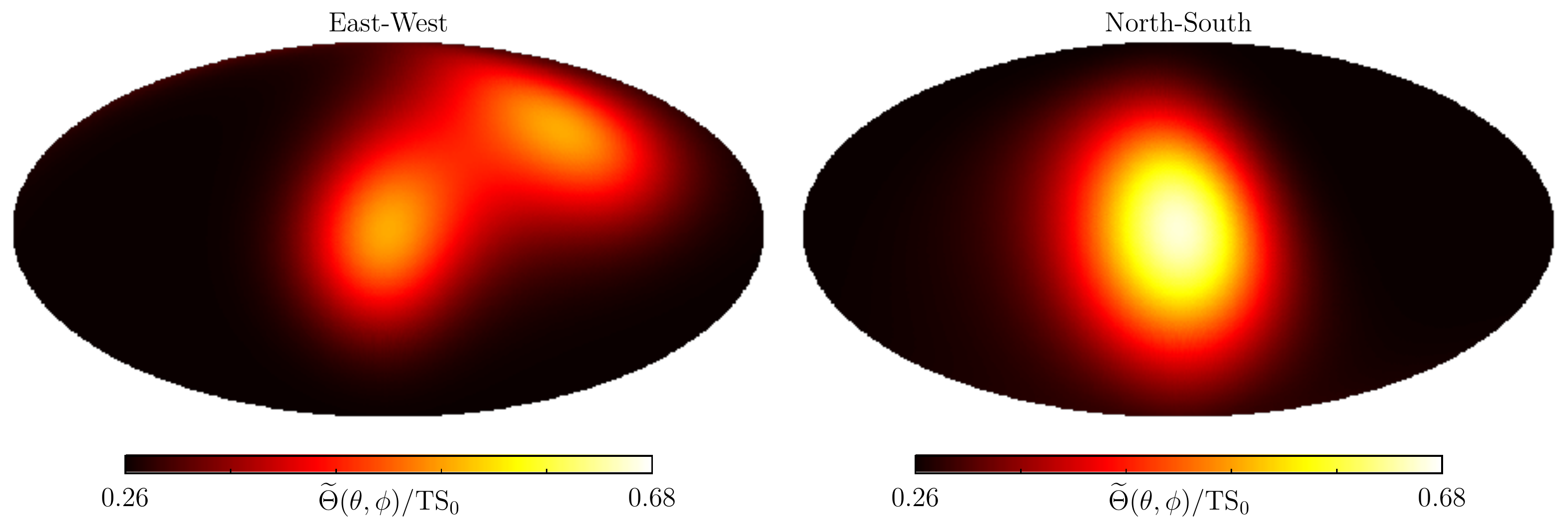}
\vspace{-0.3cm}
\caption{As in Fig.~\ref{fig:toymap}, we construct Mollweide projections of the Asimov test statistic $\widetilde \Theta(\theta, \phi)$ for the SHM rescaled by the co-located detection significance $\mathrm{TS}_0$.
However, we now perform a joint likelihood over data collected over a 24-hour period so that the daily modulation of the detector displacement vector produces a time-varying signal, which helps break degeneracies in the reconstructed directional parameters.
The Mollweide projection for a configuration in which the detectors are oriented along an East-West (North-South) orientation is shown on the left (right).
While the results of obtained in an East-West configuration do not depend on the latitude of the detectors, the North-South configuration results do, so for definiteness, we have taken the detectors to be located in New Haven, CT, the site of the HAYSTAC detector.
In both configurations, the SHM boost velocity direction can be localized effectively, although there remains a non-trivial degeneracy in the East-West map between two points on the sphere.}
\label{fig:SHM_Modulation}
\end{figure*}

While the alteration to our formalism imposed by daily modulation is minimal -- as exhibited in \eqref{eq:likelihood-daily} -- the impact on the results can be dramatic.
We will demonstrate this with several examples in this section, all within the Asimov formalism.
To begin with, we consider using $\mcn = 2$ detectors in order to determine the direction of $\vv_\odot$ in the SHM.
This is the same problem we considered in Sec.~\ref{sec:SHM}, which produced the results shown in Fig.~\ref{fig:toymap}, where there is a clear degeneracy associated with rotations around $\xx_{12}$.
We will now see explicitly that daily modulation helps lift this degeneracy.
To do so, let us suppose that the DM velocity distribution follows the SHM in \eqref{eq:SHM_full}, with $v_0 = 220$ km/s and $v_{\odot} = 232$ km/s.
Our goal, as previously, will be to infer the direction of $\hat{\vv}_{\odot}$.
We consider two detectors separated by $d = 2 \lc = 2/(m_a v_0)$, and for definiteness we place one detector at a latitude $41.3^{\circ}$ N and longitude $72.9^{\circ}$ W.
In Fig.~\ref{fig:SHM_Modulation} we show results where a second detector is placed a distance $d$ to the East (left) or North (right) of this detector with data stacked at two-hour intervals over the 24-hour period.\footnote{Note that since the Earth's rotation is aligned with the East-West direction, results obtained for the East-West configuration are independent of the exact experimental locations, so long as the detector separation is much smaller than the Earth's radius of curvature.
For any other configuration, however, the result will generically depend on latitude.}
For the North-South configuration, we see that the direction can be well-localized: a high significance axion detection in this case would lead to a precise estimation of the direction of $\vv_\odot$, as we show explicitly in Sec.~\ref{sec:MCdaily} below.
This configuration clearly outperforms an East-West configuration, where there remains a degeneracy that has not been fully lifted by the daily modulation. Additionally, the maximum test statistic realized in the North-South configuration would be approximately 10\% larger than one realized in an East-West configuration for otherwise identical data collections.

\begin{figure*}[htb]
\centering	
\includegraphics[width=.99\textwidth]{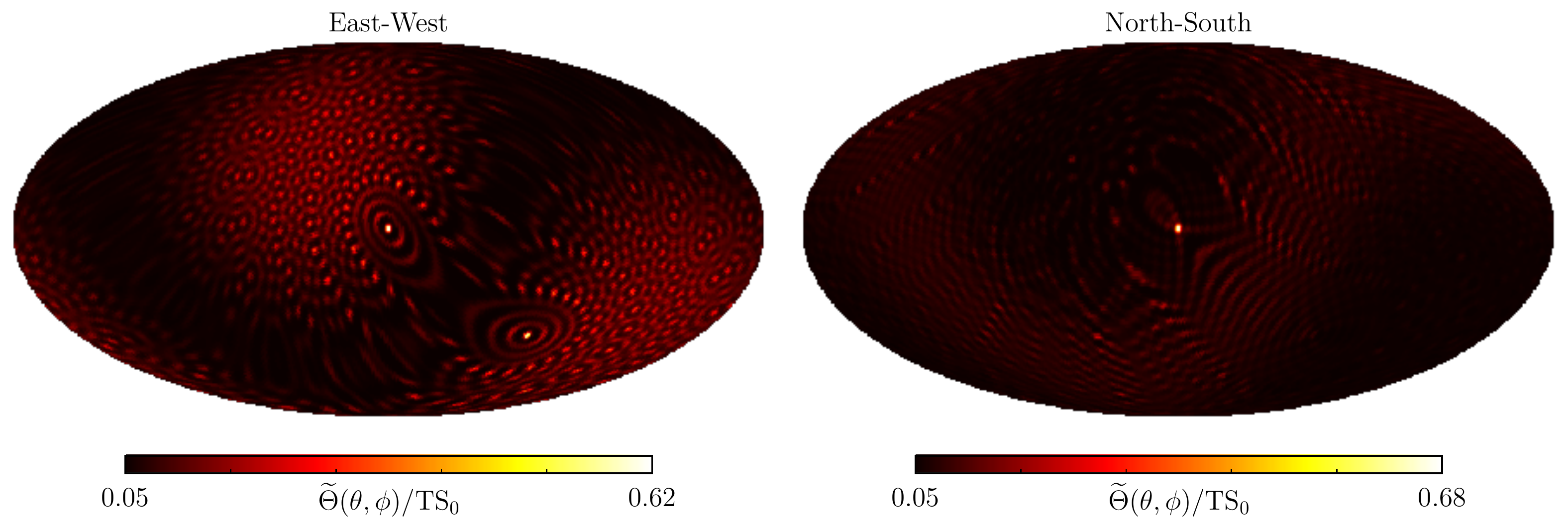}
\vspace{-0.3cm}
\caption{As in Fig.~\ref{fig:SHM_Modulation}, but for the Sagittarius stream example.
For a fixed axion mass, the physical detector separation $d = 2 \lc$ is a factor of 20 larger than in Fig.~\ref{fig:SHM_Modulation} because of the larger coherence length of the stream.
While there are many local maxima in both configurations, the North-South orientation produces only a single global maximum, at the true detector localization, while the East-West orientation leads to two degenerate global maxima (one at the true detector location and the other displaced). An animated version of these figures, showing how the localization improves throughout the day as more orientations of $\xx_{12}$ are sampled, can be found at \href{https://github.com/joshwfoster/DM_Interferometry}{github.com/joshwfoster/DM\_Interferometry}.
}
\label{fig:SGR_Modulation}
\end{figure*}

Using the same experimental design, we can also revisit the example of the Sagittarius stream discussed in Sec.~\ref{sec:stream}.
In Fig.~\ref{fig:SGR_Modulation} we construct the analogue of Fig.~\ref{fig:SHM_Modulation}, but now for the much colder stream. Note that since $v_0$ for the stream is a factor of $\sim 20$ smaller than for the SHM, the optimal detector distance $d = 2 \lc$ is a factor of 20 larger than in Fig.~\ref{fig:SHM_Modulation}. 
Although in both configurations the TS is maximized at the expected location on the sphere, nontrivial structure due to the presence of many local maxima are apparent in both the North-South and East-West configurations.
We note that, as in the SHM example, there is only one global maximum for the North-South configuration, located at the true direction of the stream. However, there remains a degeneracy in the East-West configuration.

The degeneracy represented in the Mollweide maps for the SHM and the Sagittarius stream in the East-West configuration is exact.
It has its origin in the dimensionality of the space swept out by the detector separation vector $\xx_{12}$ over the course of the day.
As studied in Sec.~\ref{sec:parameters}, for data taken at fixed $\xx_{12}$, the test statistic $\widetilde \Theta(\hat{\vv})$ evaluated as a function of the orientation of the boost velocity depends only on the angle between $\hat{\vv}$ and $\xx_{12}$.
As a result, $\widetilde \Theta(\hat{\vv}^t) = \widetilde \Theta(\hat{\vv}')$ where $\hat{\vv}^t$ is the true boost direction and $\hat{\vv}'$ is a velocity obtained by reflecting $\hat{\vv}^t$ across any plane which contains $\xx_{12}$. For detectors in an East-West configuration, the Earth's rotation produces a daily modulation of $\xx_{12}$ that is confined to the plane orthogonal to the Earth's rotational velocity vector.
As a result, the TS measured at each point in the day, and therefore the sum of such TSs, will be exactly preserved under reflections of the boost velocity across that plane.
This means that accounting for daily modulation in the East-West configuration the directional parameters can only be determined up to a reflection across the plane perpendicular to the Earth's rotation axis.
By contrast, for detectors in the North-South configuration, the set of detector separation vectors throughout the day will generically not be co-planar, and thus there is no analogous degeneracy.\footnote{An exception occurs if the two detectors have the same longitude and equal and opposite latitudes ({\it i.e.}, opposite sides of the equator on the same line of longitude).
An extreme example would be having one detector at each pole.
Then, $\xx_{12}$ is parallel to the rotation axis of the Earth and does not change direction throughout the day.
Consequently, daily modulation provides no additional information, and the full degeneracy that was present throughout Sec.~\ref{sec:parameters} returns.}

\begin{figure*}[htb]
\centering	
\includegraphics[scale = .5]{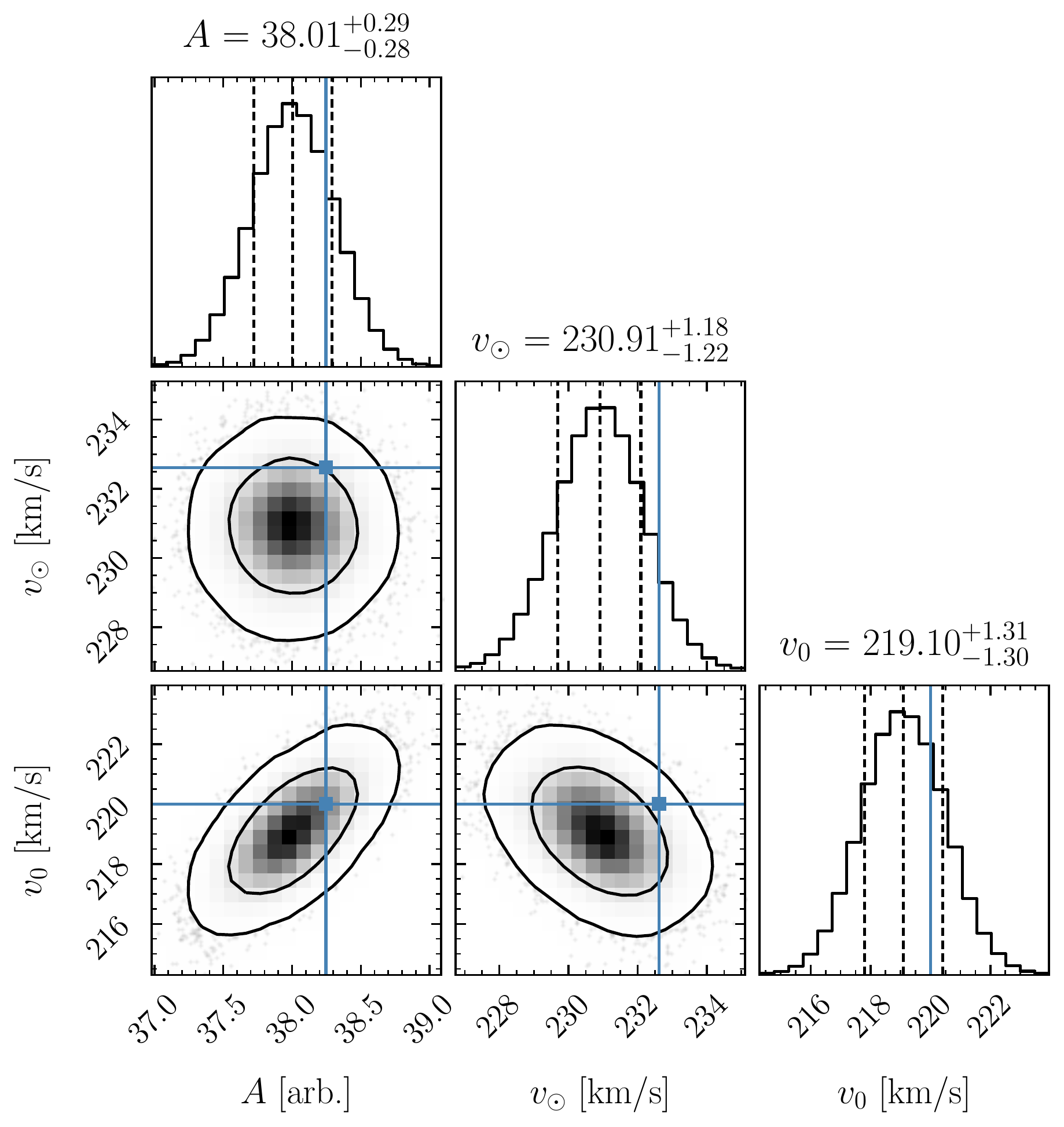}
\hspace{0.5cm}
\includegraphics[scale = .5]{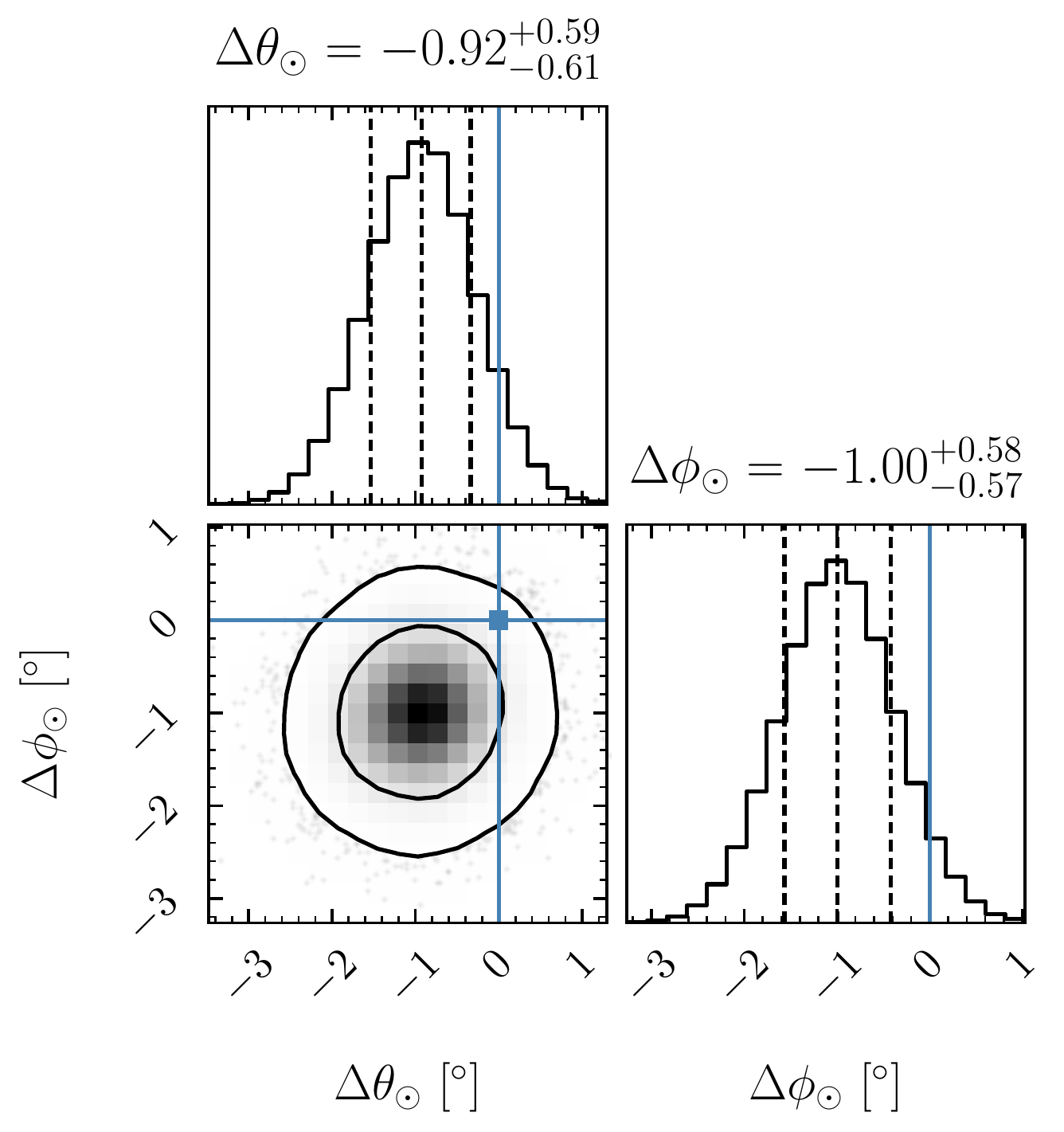}
\vspace{-0.3cm}
\caption{The posterior distribution for a model with daily modulation where the signal strength is at the threshold of an expected $5\sigma$ detection for a 100 second observation with a single detector.
Monte Carlo data is generated for 24 hours of data collection with two detectors separated along the North-South direction by a distance with $2 \times (m_a v_0)^{-1}$.
The true parameters are indicated in blue, with the 1$\sigma$ confidence intervals on the parameter estimations are indicated by the dashed black lines in the single-parameter posteriors.
The two parameter posteriors show the $1\sigma$ and $2\sigma$ contours.
On the left, we display the posterior distributions for the overall signal strength, the boost speed of the SHM, and the velocity dispersion of the SHM, all of which are parameters accessible in a single detector configuration.
On the right, we display the posterior distributions for the angles $\Delta \theta_\odot = \theta_\odot - \theta^t_\odot$ and $\Delta \phi_\odot = \phi_\odot - \phi^t_\odot$ which specify the orientation of SHM boost velocity and are only accessible in a multiple-detector configuration.
Both $\theta_\odot$ and $\phi_\odot$ are determined at degree precision in this scenario.}
\label{fig:SHM_Multinest}
\end{figure*}

%%%%%%%%%%%%%%%%%
\subsection{Monte Carlo Example with Daily Modulation}
\label{sec:MCdaily}
%%%%%%%%%%%%%%%%%

As a realistic demonstration of our ability to perform parameter estimation using the daily modulation effect, we generate a Monte Carlo realization of data in the North-South SHM scenario (as depicted in the right panel of Fig.~\ref{fig:SHM_Modulation}) using $A= 38.25$ and $\lambda_B = 1$, both of which we take to be dimensionless without loss of generality.
The values of $A$ and $m_a$ was chosen to generate a signal of expected $5\sigma$ significance during a 100-second collection in a single detector to mimic a realistic resonant scanning strategy in which one of two independently operated detectors detects an excess and both are then used for a 24-hour observation of the excess candidate. We constructed 24 hours of Monte Carlo data for this signal, taking a detector separation of $d = 2 \lc$; with these parameters, the excess would be expected to appear at ${\rm TS}_0 \approx 60,000$ after 24 hours.
While large, this ${\rm TS}$ is consistent with the power of a resonant strategy once the axion mass is known.

Using uniform priors on $A$ between $[33, 43]$, on $v_\odot$ between $[212.5, 252.5]$ km/s, on $v_0$ between $[200, 240]$ km/s, on $\lambda_B$ between $[.999, 1.001]$ and a uniform prior on the sphere for $(\theta_\odot, \phi_\odot)$, we construct a Bayesian posterior distribution for the model parameters. The results of an analysis performed using Multinest~\cite{Feroz:2007kg, Feroz:2008xx, Feroz:2013hea, 2014A&A...564A.125B} with 2,000 live points are shown in Fig.~\ref{fig:SHM_Multinest}. 
In particular, we see that the true location of the stream has been located to degree precision.
This precision can be understood from \eqref{eq:TSv0scale}, which gives the expectation $\sigma_{\theta_\odot} \sim 0.5^\circ$, consistent with what is shown in the figure. Let us suppose that the Sagittarius stream, as modeled in this work, comprises 10\% of the local DM.
In the example above, we would expect that after 24 hours the location of the stream could be localized to $\sim10'$;  interestingly, this represents greater accuracy for stream localization than localization of the bulk SHM even though the stream is a sub-dominant component of the DM.

%%%%%%%%%%%%%%%%%
\section{Conclusion}
\label{sec:conclusions}
%%%%%%%%%%%%%%%%%

In this work we have demonstrated the power of DM interferometry for wave-like DM.
The spatial coherence of the DM field imprints phase correlations on the signals observed at spatially-separated detectors, and these phase correlations are sensitive to parameters in the full 3-dimensional velocity distribution $f(\vv)$, whereas a single detector is blind to all effects beyond the speed distribution $f(v)$.
As a result, the advantages of DM interferometry go beyond a simple coherent enhancement of the signal strength as the number of detectors is increased.
By taking advantage of the fact that the correlation matrix of the Fourier-transformed signals at multiple detectors depends on modified speed distributions which contain modulated forms of $f(\vv)$, we have demonstrated that parameters such as the solar velocity vector may be reliably extracted from two detectors separated by a distance $d \sim \lc$.
Furthermore, directional parameters of coherent substructure such as DM streams may be estimated at even higher significance, though in that case the optimal separation $\lc$ is parametrically different from the DM de Broglie wavelength $\ldB$.

Our formalism has immediate practical applications for new and upcoming axion DM experiments.
The sensitivity to $\gagg$ for resonant-cavity axion experiments which use external magnetic fields, like ADMX and HAYSTAC, is typically $BV^{1/2}$, where $B$ is the peak magnetic field strength and $V$ is the magnetic field volume.
In order to achieve resonant enhancement, the volume of an individual cavity is fixed to be of order $1/m_a^3$, so to achieve greater sensitivity, one must either increase the $B$-field or construct a multiplexed readout with multiple cavities.
Assuming the latter strategy is chosen, our results motivate placing at least one of the cavities at a distance $\lc$: if a signal is detected, the loss of coherent enhancement of the signal is more than compensated by the ability to localize the boost direction of the DM velocity distribution to within 1 degree with just 24 hours of data.

While there are many challenges to the construction of additional instruments, we emphasize that all of the important phenomenology is captured by a two-detector array. This smoking-gun signature of DM is invisible to a multiplexed setup where all cavities lie inside a single coherence length.
A similar analysis applies to experiments in the quasistatic regime like ABRACADABRA and DM-Radio, where the physical volume of the experiment is decoupled from $m_a$. 
For both types of experiments, our formalism may be applied to determine the optimal detector orientation for localizing the solar velocity to the desired precision (with North-South orientations generally being preferred to East-West). The optimal detector separation corresponds to physically reasonable distances for well-motivated axion masses -- $\mathcal{O}(10)$ m for the $10^{-5}$ eV mass range of ADMX and HAYSTAC and $\mathcal{O}(1000)$ km for the $10^{-9}$ eV targeted by ABRACADABRA/DM-Radio -- and as such the coherence length and detector orientation can form an important design parameter for future experiments, in much the same way as $L/E$ determines the design of neutrino oscillation experiments.

The future of axion detection involves readout beyond the standard quantum limit, using tools such as Josephson parametric amplifiers and squeezed states.
In this regime, it is important to note that our variables $R_k$ and $I_k$ are canonically conjugate, and thus cannot be simultaneously measured to arbitrary precision.
In future work, we plan to investigate how our formalism must be modified for quantum-limited readouts. As the number of new axion experiments proliferates, this work motivates careful consideration of the spatial configuration of multiplexed detectors.

%%%%%%%%%%%%%%%%%
\section*{Acknowledgments}
%%%%%%%%%%%%%%%%%

We thank members of the ABRACADABRA Collaboration (Andrew Gavin, Reyco Henning, Jonathan Ouellet, Kaliro\"{e} Pappas, Chiara Salemi, and Lindley Winslow), as well as Kelly Backes, Karl Van Bibber, Aaron Chou, and Andrei Derevianko, for enlightening discussions.
JF and BRS were supported in  part  by  the  DOE  Early Career  Grant  DESC0019225.
The work of YK is supported in part by US Department of Energy grant DE-SC0015655.
RN is supported by an NSF Graduate Fellowship.
NLR is supported by the Miller Institute for Basic Research in Science at the University of California, Berkeley.
This work used computational resources  and  services  provided  by  Advanced  Research Computing  at the  University  of  Michigan,  Ann  Arbor, in addition to the Lawrencium computational cluster resource provided by the IT Division at the Lawrence Berkeley National Laboratory supported by the Director, Office of Science, and Office of Basic Energy Sciences, of the U.S. Department of Energy under Contract No. DE-AC02-05CH11231.

\appendix
%%%%%%%%%%%%%%%%%
\section{Coherence Length and Time}
\label{app:coherence}
%%%%%%%%%%%%%%%%%

In this appendix we briefly review the concepts of the coherence length and time, as relevant to wavelike DM.
We emphasize that in our work both concepts only arise heuristically.
Indeed, the coherence length and time are only defined parametrically, and for all quantitative results we instead rely on the likelihood formalism in~\cite{Foster:2017hbq}, which produces not only all parametric scalings, but also the required $\mathcal{O}(1)$ factors.

Consider first the coherence length $\lc \sim (\mDM v_0)^{-1}$, the scale over which wavelike DM remains coherent.
In discussions of ultralight DM, ``coherence length'' is often used interchangeably with ``de Broglie wavelength.''
Strictly speaking, though, the de Broglie wavelength $\ldB = 2\pi/(m_a v)$ is a property of particles with fixed velocity $v$, while the coherence length describes the dephasing of various plane wave components with different velocities.
When $v_0 \sim v$, these two length scales are comparable, but there are relevant situations where the two diverge, such as for cold streams, and then the distinction between the coherence and de Broglie wavelengths becomes important.

The coherence time is then the timescale over which a measured signal of ultralight DM will build up coherently.
In real space, this is the time it take for a new spatially coherent packet of the DM wave, which has size $\lc$, to arrive at the instrument.
If these packets travel with a mean speed of $\bar{v}$, then the timescale is $\tau \sim \lc/\bar{v} \sim (\mDM \bar{v}\, v_0)^{-1}$.
The same result can be arrived at from a frequency space consideration.
The Fourier transform of an experimental data set collected over a time $T$ will have a frequency resolution of $\Delta \omega = 2\pi/T$.
If the entire signal fits within a single frequency bin, the result is associated with a single draw from an exponential distribution, as shown in~\cite{Foster:2017hbq}.
Once we resolve the signal, however, we obtain multiple draws which will combine incoherently, partially offsetting the benefit of additional integration time.
The coherence time is therefore dictated by the width of the signal in frequency space, and then as $d \omega = \mDM v\, dv$, we again arrive at $\tau \sim (\mDM \bar{v}\, v_0)^{-1}$.

%%%%%%%%%%%%%%%%%
\section{Demonstrating $\dd \sim \mcn(\mathbf{0},\mathbf{\Sigma})$}
\label{app:derivation}
%%%%%%%%%%%%%%%%%

The goal of this appendix is to demonstrate a fact that was used without proof in the main body: the data set $\dd$, given in \eqref{eq:data_vector}, is a random variable drawn from a multivariate normal distribution with zero mean and covariance matrix $\bms$ as given by \eqref{eq:fullcovariance}.
In order to show this we will start from the known statistics of the axion field, as reviewed in the main body, together with a Gaussian background, and show that the mean and variances of the data sets follow the expected normal distribution.
We will further confirm this result with a Monte Carlo realization of the axion field.
From here, rather than confirm that all higher moments are also consistent with Gaussianity, we will instead confirm numerically that the distribution is normal.
Indeed, the diagonal components of $\dd$, which govern the statistics of individual detectors, must be Gaussian as proven in~\cite{Foster:2017hbq}.

Let us begin by restating \eqref{eq:axionfield} in a simplified notation.
We introduce a single multi-index $d=abc$, and a random variable $f_{d} = \alpha_{d} \sqrt{f(\vv_{d}) (\Delta v)^3}$, yielding
\es{}{
a(\xx,t)
= \frac{\sqrt{\rhoDM}}{m_a} \sum_{d} f_d \cos\left( \omega_d t - m_a \vv_d \cdot \xx + \phi_d \right).
}
We now envision collecting a data sensitive to this axion field at each of the $\mcn$ detectors, located at positions $\xx_i$.
Specifically, we imagine collecting $N$ measurements at a frequency $f = 1/\Delta t$ at each experiment, so that we have at our disposal $N \times \mcn$ data points $\{\Phi_n^{(i)}\}$, with
\es{}{
\Phi_n^{(i)} = m_a \sqrt{\frac{A_i}{\rhoDM}} a_n(\xx_i,\,n \Delta t) + x^{(i)}_n\,.
}
The second term in this expression captures the background noise.
We will assume the noise is Gaussian, which holds for a wide range of sources as described in the main body, and in detail that it satisfies
\es{eq:bkg-stat}{
\left\langle x^{(i)}_n \right\rangle = 0\,,\hspace{0.5cm}
\left\langle x^{(i)}_n x^{(j)}_m \right\rangle = \delta_{ij} \delta_{nm} \frac{\lambda_{B,i}}{\Delta t}\,.
}
In other words, we assume the noise has zero mean, is uncorrelated between detectors, and has a variance that increases with the measurement frequency $f$.
The variance is controlled by the mean power in the background, $\lambda_{B,i}$, and if there are multiple background sources at a single detector, their power can simply be combined.

From this data set, we compute the discrete Fourier transform $\{\Phi_k^{(i)}\}$ using \eqref{eq:DFT}, and then the associated real and imaginary parts, $R_k^{(i)}$ and $I_k^{(i)}$ from \eqref{eq:re-im}.
These variables are what combine to form the data vector $\dd$, and so the goal is to study their statistics.
Before proceeding, let's introduce some further notation to keep expressions compact.  
Firstly, we encapsulate the axion phase into a single term,
\es{}{
\varphi_{d,n}^{(i)} = \omega_d n \Delta t - m_a \vv_d \cdot \xx_i + \phi_d\,.
}
To capture the trigonometric sums introduced by the Fourier transforms, we write
\es{}{
c_{n,k} = \cos \left( \frac{2\pi k n}{N} \right) = \cos \left( \omega n \Delta t \right),
}
and the equivalent expression for sine is denoted $s_{n,k}$.
Using this, the real and imaginary parts of the data set can be written
\begin{align}
R_k^{(i)} = &\frac{\Delta t}{\sqrt{T}}\sum_{n=0}^{N-1} \left[ \sqrt{A_i} \sum_{d} f_d \cos\varphi_{d,n}^{(i)} + x^{(i)}_n \right] c_{n,k}, \\
I_k^{(i)} = &- \frac{\Delta t}{\sqrt{T}}\sum_{n=0}^{N-1} \left[ \sqrt{A_i} \sum_{d} f_d \cos \varphi_{d,n}^{(i)} + x^{(i)}_n \right] s_{n,k}\,. \nonumber
\end{align}
From these expressions, we can see immediately that $\langle R_k^{(i)} \rangle = \langle I_k^{(i)} \rangle = 0$.
That this holds for the background follows from \eqref{eq:bkg-stat}, and for the axion signal contribution we have
\es{}{
\big\langle f_d \cos \varphi_{d,n}^{(i)} \big\rangle
= \langle f_d \rangle \big\langle \cos \varphi_{d,n}^{(i)} \big\rangle = 0\,.
}
The first step follows as the value of $\alpha_d$ (and hence $f_d$) is uncorrelated with $\phi_d$ (and hence $\varphi_{d,n}^{(i)})$, whilst the second utilizes the fact $\langle \cos \varphi \rangle = 0$ when the argument $\varphi$ is a random phase.
This establishes that $\langle \dd \rangle = {\bf 0}$.

Next we consider the covariances.
In particular, we will compute $\langle R_k^{(i)} R_k^{(j)} \rangle$.
The calculation where one or both of the real components is replaced by an imaginary equivalent proceeds similarly, and we will comment on the important differences throughout.
In detail, we will compute
\begin{align}
&\left\langle R_k^{(i)} R_k^{(j)} \right\rangle = \frac{(\Delta t)^2}{T} \nonumber\\
\times&\left \langle \sum_{n=0}^{N-1} \left[ \sqrt{A_i}\sum_d f_d \cos \varphi_{d,n}^{(i)} + x^{(i)}_n \right] c_{n,k} \right.\\
\times&\left.\sum_{m=0}^{N-1} \left[ \sqrt{A_j}\sum_s f_s \cos\varphi_{s,m}^{(j)} + x^{(j)}_m \right] c_{m,k} \right \rangle\,. \nonumber
\end{align}
Note the effect of sending $R_k^{(j)} \to I_k^{(j)}$ is simply to replace $c_{m,k} \to -s_{m,k}$, and similarly for $R_k^{(i)}$.
Continuing with the calculation at hand, expanding out the final two lines, we will have expressions involving only the signal, only the background, and cross terms.
As the background value is uncorrelated with the signal, the cross terms will be zero via an almost identical argument to the vanishing of the means.
Of the remaining terms, consider the background first.
\es{}{
&\frac{(\Delta t)^2}{T} \left \langle \sum_{n,m=0}^{N-1} \left( x^{(i)}_n c_{n,k} \right) \left(  x^{(j)}_m c_{m,k} \right) \right \rangle \\
= &\frac{\delta_{ij} \lambda_{B,i}}{N} \sum_{n=0}^{N-1} (c_{n,k})^2
= \frac{\delta_{ij} \lambda_{B,i}}{2}\,,
}
which holds except for $k=0$ (or $k=N/2$ for $N$ even).
Note if we were evaluating $\langle I_k^{(i)} I_k^{(j)} \rangle$, we would have the same expression but with $c_{n,k} \to s_{n,k}$, and therefore the background contribution would be identical.
If we were evaluating $\langle R_k^{(i)} I_k^{(j)} \rangle$, however, the background contribution would vanish as $\sum c_{n,k} s_{n,k} = 0$.
Taken together, these results demonstrate the appearance of $\lambda_B$ in \eqref{eq:fullcovariance}.

\begin{figure*}[htb]
\centering	
\includegraphics[width=.47\textwidth]{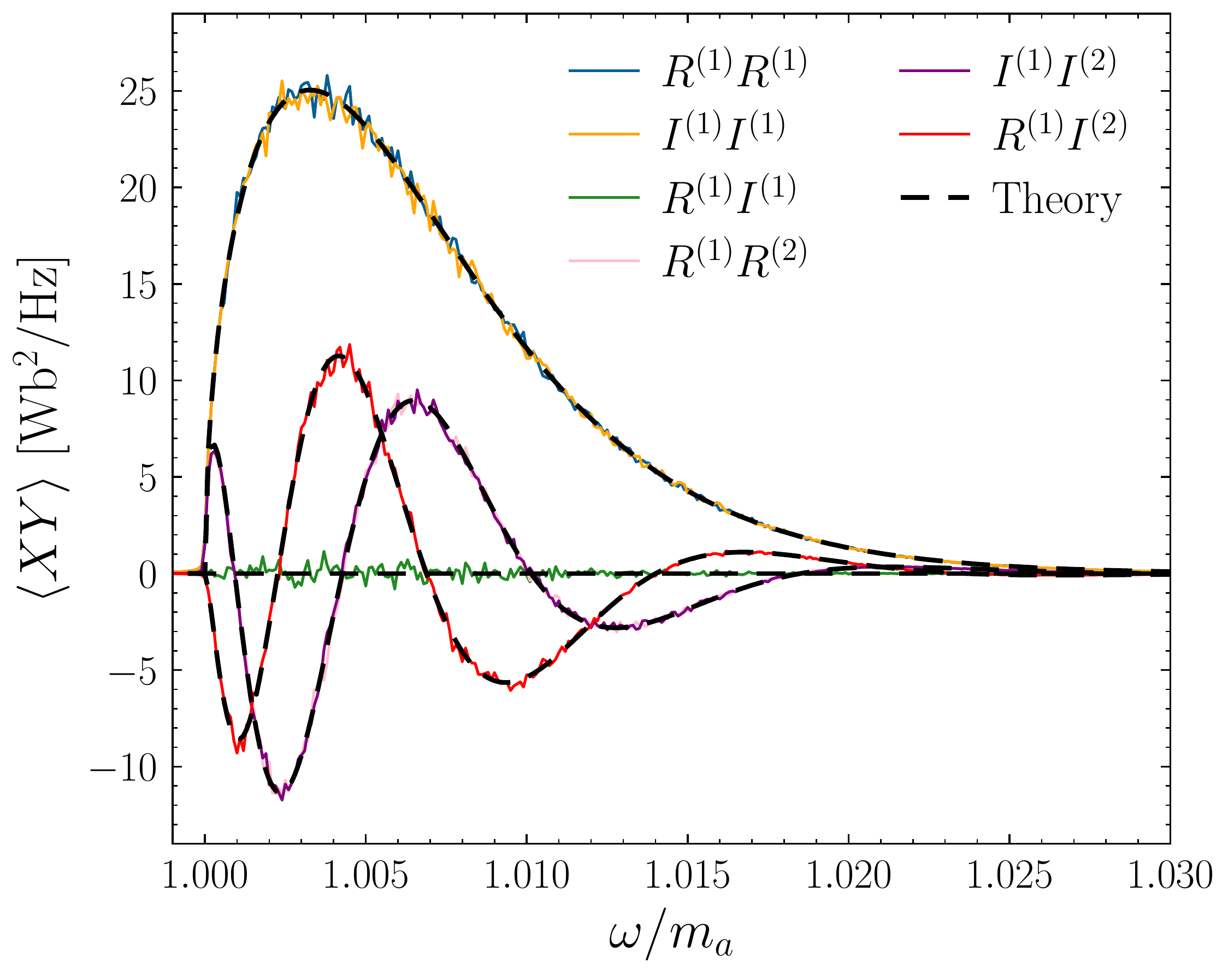}
\hspace{0.2cm}
\includegraphics[width=.47\textwidth]{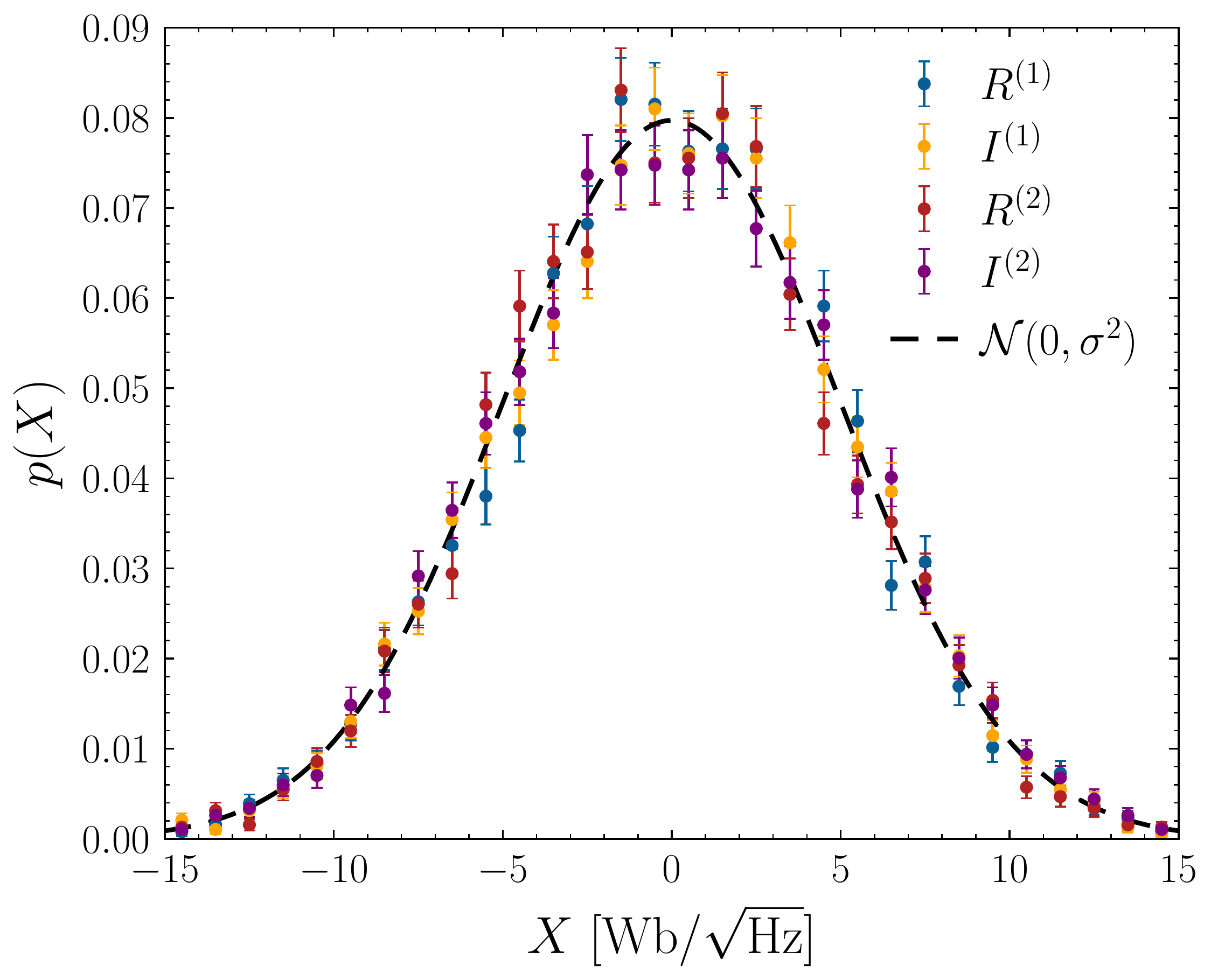}
\vspace{-0.3cm}
\caption{Monte Carlo validation that the statistics of DM interferometry are as claimed in App.~\ref{app:derivation}.
In the left figure we confirm that the variances of the real and imaginary signal-only data sets, collected for the $\mcn=2$ experiments, is as claimed in \eqref{eq:fullcovariance}.
This was proven directly in the text, but in the plot we show that the average of 4,000 Monte Carlo simulations provides a consistent prediction for the variances as a function of frequency in the different cases.
On the right figure, for the frequency where $\langle R^{(1)} R^{(1)} \rangle$ achieves its maximum, we show the distribution of values across the simulations.
In detail, we see that the real and imaginary components are normally distributed, and consistent with a mean-zero normal distribution, where the variance is given as on the left, here $\sigma^2 \approx 25~{\rm Wb^2/Hz}$.
We found that the distributions were consistent with the Gaussian expectation at the level of $p > .05$ using the D'Agostino and Pearson omnibus normality test~\cite{10.2307/2334522, 10.1093/biomet/60.3.613}.
In both cases, each Monte Carlo simulation involves a direct construction of the axion field starting from \eqref{eq:aplanewaves} with $N_a=100,000$, taking $m_a = 2\pi~{\rm Hz}$, and $A = 1~{\rm Wb}^2$.
Further, we take the velocity distribution to follow a variant of the SHM in \eqref{eq:SHM_full}, but with $v_0 = 0.07$ and $\vv_{\odot} = (0,0.08,0)$, both in natural units. The (unphysically) large velocity helps simplify the computation of the Fourier transform.
The detector separation is $\xx_{12} = d (0, 1, 0)$, with $d \approx 4.4 \lc$.
}
\label{fig:MC-validation}
\end{figure*}

Now we turn to the signal contribution, for the moment dropping the overall factor of $\sqrt{A_i A_j} (\Delta t/N)$,
\es{}{
&\left \langle \sum_{n,m=0}^{N-1} \sum_{d,s} f_d f_s \cos \varphi_{d,n}^{(i)} \cos\varphi_{s,m}^{(j)} c_{n,k} c_{m,k} \right \rangle \\
= &\sum_{n,m=0}^{N-1} c_{n,k} c_{m,k} \sum_{d,s} \langle f_d f_s \rangle \langle \cos \varphi_{d,n}^{(i)} \cos\varphi_{s,m}^{(j)} \rangle\,.
}
Again we used the independence of the amplitude and phase of the random walk that emerges in calculating the axion field statistics.
The second expectation value in this expression will vanish unless the random phases in the cosines are equal, effectively as
\es{}{
\langle e^{i(\phi_{d}-\phi_{s})} \rangle = \delta_d^s\,.
}
Further, as $\langle \alpha_d^2 \rangle = 2$, we can also evaluate the result as
\es{}{
&\sum_{d,s} \langle f_d f_s \rangle \langle \cos \varphi_{d,n}^{(i)} \cos\varphi_{s,m}^{(j)} \rangle \\
= &2 \sum_d f(\vv_d) (\Delta v)^3 \langle \cos \varphi_{d,n}^{(i)} \cos\varphi_{d,m}^{(j)} \rangle\,,
}
which we can simplify further as,
\begin{align}
\label{eq:RRsig5}
&\langle \cos \varphi_{d,n}^{(i)} \cos\varphi_{d,m}^{(j)} \rangle\\
= &\frac{1}{2} \left[
\cos\left( \omega_d (n-m) \Delta t - m_a \vv_d \cdot \xx_{ij} \right) \right. \nonumber \\
+&\left. \left \langle \cos\left( \omega_d (n+m) \Delta t - m_a \vv_d \cdot (\xx_i+\xx_j) + 2\phi_d \right) \right \rangle \right] \nonumber\\
= &\frac{1}{2} \cos\left( \omega_d (n-m) \Delta t - m_a \vv_d \cdot \xx_{ij} \right) \nonumber \\
= &\frac{1}{2} \cos\left( \frac{\omega_d}{\omega} \frac{2\pi k(n-m)}{N} \right) \cos \left( m_a \vv_d \cdot \xx_{ij} \right) \nonumber\\
+ &\frac{1}{2} \sin\left( \frac{\omega_d}{\omega} \frac{2\pi k(n-m)}{N} \right) \sin \left( m_a \vv_d \cdot \xx_{ij } \right). \nonumber
\end{align}
In the final step we see the emergence of the $\kk \cdot \xx$ type phase factors that separate $\FF^{c,s}_{ij}(v)$ defined in \eqref{eq:F-c-s} from $f(v)$.
We have broken the calculation into a number of pieces at this stage, let us begin to put things back together.
Combining the different expressions above, we have
\begin{widetext}
\begin{align}
\left\langle R_k^{(i)} R_k^{(j)} \right\rangle 
= &\,\frac{1}{2} \delta_{ij} \lambda_{B,i}(\omega)
+ \sqrt{A_i A_j} \frac{\Delta t}{N} \sum_d f(\vv_d) (\Delta v)^3
\left[ \cos \left( m_a \vv_d \cdot \xx_{ij} \right)  \sum_{n,m=0}^{N-1} c_{n,k} c_{m,k} \cos\left( \frac{\omega_d}{\omega} \frac{2\pi k(n-m)}{N} \right) \right. \nonumber\\
&\hspace{6.25cm}\left.+ \sin \left( m_a \vv_d \cdot \xx_{ij } \right) \sum_{n,m=0}^{N-1} c_{n,k} c_{m,k} \sin\left( \frac{\omega_d}{\omega} \frac{2\pi k(n-m)}{N} \right) \right] \nonumber\\
= &\,\frac{1}{2} \delta_{ij} \lambda_{B,i}(\omega)
+ \frac{\pi\sqrt{A_i A_j}}{2} \sum_d (\Delta v)^3 f(\vv_d)
\cos \left( m_a \vv_d \cdot \xx_{ij} \right) \delta(\omega_d-\omega) \\
= &\,\frac{1}{2} \delta_{ij} \lambda_{B,i}(\omega)
+ \frac{\pi\sqrt{A_i A_j}}{2 m_a \vw} \int d^3 \vv f(\vv)
\cos \left( m_a \vv \cdot \xx_{ij} \right) \delta(|\vv|-\vw) \nonumber \\
= &\,\frac{1}{2} \left[ c_{ij}(\omega) + \delta_{ij} \lambda_{B,i}(\omega) \right]. \nonumber
\end{align}
\end{widetext}
The final result is the claimed form of $\langle R_k^{(i)} R_k^{(j)} \rangle$ used in the main body, but let us detail the steps in the calculation, working backwards.
In the last step we simply recalled the definitions introduced in \eqref{eq:F-c-s} and \eqref{eq:c-s}.
The penultimate step simply involved approximating the sum over all velocity components $d=abc$ with an equivalent integral.
The only non-trivial manipulation occurred when we evaluated the sums over $n$ and $m$.
These were performed using a set of discrete Fourier transform double orthogonality relations, which for convenience we have collected in App.~\ref{app:orthogonality}.
From those relations, we can see that as $\langle R_k^{(i)} R_k^{(j)} \rangle$ involved $c_{n,k} c_{m,k}$, only the cosine of $\kk \cdot \xx_{ij}$ survived.
By analogy, if we were evaluating $\langle I_k^{(i)} I_k^{(j)} \rangle$, we would instead have $s_{n,k} s_{m,k}$ in the sums, which would again isolate the cosine.
On the other hand, for $\langle R_k^{(i)} I_k^{(j)} \rangle$ (where the background contribution vanishes as described above), we have $c_{n,k} s_{m,k}$, which instead singles out the sine, implying the above result would have $c_{ij}(\omega) \to s_{ij}(\omega)$.
The same argument holds for $\langle I_k^{(i)} R_k^{(j)} \rangle$, up to a sign.

Taken together, the above arguments suffice to demonstrate analytically that the variance of the data set is as claimed in the main body.
We can also confirm this result numerically.
On the left of Fig.~\ref{fig:MC-validation}, we show that a direct construction of the axion field as a sum over $N_a$ plane wave components,
\es{eq:aplanewaves}{
a(\xx, t) = \frac{\sqrt{2\rhoDM}}{m_a\sqrt{N_a}}\sum_{i =1}^{N_a} \cos \left [\omega_i t - m_a\vv_i \cdot \xx + \phi_i \right],
}
where $\vv_i$ is drawn from $f(\vv)$ and $\phi_i$ is drawn uniformly from $[0, 2\pi)$, leads to the exact same results.\footnote{Binning the velocities leads to \eqref{eq:axionfield} in the main text, with a Rayleigh-distributed amplitude in each bin.}
The detailed parameter choices are described in the figure caption, and the curves represent the average over repeating this procedure 4,000 times.
In all cases, there is excellent agreement between this approach and the corresponding theory curves.

On the right of Fig.~\ref{fig:MC-validation} we confirm a point that we did not demonstrate directly, namely that the individual real and imaginary components are normally distributed.
The distribution is shown amongst the 4,000 simulated data sets for the two components measured at two different detectors.
In all cases consistency is observed with the predicted Gaussian distribution.
We performed a chi-squared test to determine the goodness of fit and found $p$-values greater than $0.05$.
In detail, the $R^{(1)}$, $I^{(1)}$, $R^{(2)}$, and $I^{(2)}$ data sets shown in Fig.~\ref{fig:MC-validation}, had corresponding $p$-values of $0.06$, $0.12$, $0.97$, and $0.27$.

%%%%%%%%%%%%%%%%%
\section{Orthogonality Relations}
\label{app:orthogonality}
%%%%%%%%%%%%%%%%%

In App.~\ref{app:derivation} we made use of several unstated orthogonality relations.
We collect these in the present appendix.
Firstly, the following expressions vanish for any $k$
\es{}{
&\sum_{n,m=0}^{N-1}  c_{n,k} s_{m,k}
\cos\left( \frac{\omega_d}{\omega} \frac{2\pi k (n-m)}{N} \right) \\
= &\sum_{n,m=0}^{N-1} s_{n,k} c_{m,k}
\cos\left( \frac{\omega_d}{\omega} \frac{2\pi k (n-m)}{N} \right)\\
= &\sum_{n,m=0}^{N-1} c_{n,k} c_{m,k}
\sin\left( \frac{\omega_d}{\omega} \frac{2\pi k (n-m)}{N} \right) \\
= &\sum_{n,m=0}^{N-1}  s_{n,k} s_{m,k}
\sin\left( \frac{\omega_d}{\omega} \frac{2\pi k (n-m)}{N} \right) \\
= &\,0\,.
}
However, there are four non-zero combinations.
In detail, for most values of $k$,
\es{eq:OR-nz1}{
&\sum_{n,m=0}^{N-1} c_{n,k} c_{m,k}
\cos\left( \frac{\omega_d}{\omega} \frac{2\pi k (n-m)}{N} \right) \\
= &\sum_{n,m=0}^{N-1}  s_{n,k} s_{m,k}
\cos\left( \frac{\omega_d}{\omega} \frac{2\pi k (n-m)}{N} \right) \\
= &\sum_{n,m=0}^{N-1}  c_{n,k} s_{m,k}
\sin\left( \frac{\omega_d}{\omega} \frac{2\pi k (n-m)}{N} \right) \\
= &\sum_{n,m=0}^{N-1}  s_{n,k} c_{m,k}
\sin\left( \frac{\omega_d}{\omega} \frac{2\pi k (n-m)}{N} \right) \\
= &\,\left(\frac{N}{2} \right)^2 \frac{2\pi}{T} \delta(\omega_d-\omega).
}
The exception to the above is if $k=0$, or $k=N/2$ for $N$ even.
For those values, only one of the above three sums is non-zero, in detail
\es{eq:OR-nz2}{
&\sum_{n,m=0}^{N-1} c_{n,k} c_{m,k}
\cos\left( \frac{\omega_d}{\omega} \frac{2\pi k (n-m)}{N} \right) \\
= &\,N^2 \frac{2\pi}{T} \delta(\omega_d-\omega).
}
However, recall that we usually exclude these exceptional $k$ values from our likelihood.

The non-zero results above were written in terms of Dirac $\delta$-functions, however this is an approximation.
Recall all results are obtained through the discrete Fourier transform, within which the frequency can be interpreted as $\omega = (2 \pi/T) k$, with $k = 0,1,\ldots,N-1$.
In truth, if we define $k_d = \lfloor \omega_d T/2\pi \rfloor$, then what appears in the above sums is the Kronecker-delta $\delta_{k_d}^k$.
However, in the spirit of assuming our frequency resolution is sufficient enough to approximate $\omega$ as a continuous variable, we take
\es{}{
\delta_{k_d}^k = \delta(k_d-k) = \frac{2\pi}{T} \delta(\omega_d-\omega),
}
which is the form it appears in \eqref{eq:OR-nz1} and \eqref{eq:OR-nz2}.

%%%%%%%%%%%%%%%%%
\section{Data Stacking Procedure}
\label{app:stack}
%%%%%%%%%%%%%%%%%

In practical situations it is usually neither feasible nor necessary to save the entire time-series data to disk and then construct the Fourier transform of the full data set.
The frequency resolution of this complete Fourier transform would be $\Delta \omega = 2\pi/T$, and potentially much smaller than the scale of any expected features induced by the signal due to $f(\vv)$.
As a specific example, the ABRACADABRA-10 cm experiment~\cite{Ouellet:2018beu,Ouellet:2019tlz} recorded the PSD data over short time periods and then stacked the PSD data over the time subintervals to construct the average PSD data.
The advantage of this averaging procedure is that it requires less storage and is easier to deal with computationally, since there are less frequencies involved than would be in the full data set without time sub-binning.

With this in mind, it is useful to understand how we may stack the Fourier transform data over multiple experiments in such a way that we preserve the full power of the likelihood in~\eqref{eq:likelihood} but that allows us to reduce the data volume needed to be saved to disk.
(An optimized procedure for stacking the data from a single experiment is presented in~\cite{Foster:2017hbq}.)
Let us imagine that we record time-series data in $N_T$ equal time subintervals of time $\Delta T = T/N_T$, and that in each subinterval the frequency spacing of the $\Delta N = N/N_T$ Fourier components is sufficient to resolve the axion signal by multiple frequency bins, {\it i.e.} we retain sufficient frequency resolution that our signal remains well resolved.
We then denote the full data set by $d = \{ \dd_k^{\ell} \}$, indexed now by both $k=1,\ldots,\Delta N-1$, denoting the Fourier component, and $\ell=1,\ldots,N_T$, the data subinterval.
The appropriate likelihood is then simply the product of the likelihood in~\eqref{eq:likelihood}, but now also over all values of $N_T$.
However as $N_T \times \Delta N = N$, the number of frequency bins in the Fourier transform of the full data, at this stage we have not reduced the size or complexity of the data or likelihood evaluations at all.
In order to do so, we can combine the data into the following average data matrix, which can be computed prior to any evaluation likelihood,
\es{eq:av-data}{
[\bar{\dd}_k]^{ij} = \frac{1}{N_T} \sum_{\ell=1}^{N_T} d_{k,\ell}^i d_{k,\ell}^j\,.
}
Here, the indices $i$ and $j$ run over the $2\mcn$ entries of the data vector in \eqref{eq:data_vector}, $k$ indexes the discrete Fourier transform, and $\ell$ specifies the appropriate subintervals.
In terms of the average data matrix, the likelihood can be written as
\es{eq:likelihood-stacked}{
\mathcal{L} (d | \mathcal{M}, \bmt) = \prod_{k=1}^{\Delta N-1} 
\frac{\exp \left[ - \frac{N_T}{2} {\rm Tr} ( \bar{\dd}_k \cdot \bms^{-1}_k) \right]}{[(2 \pi)^{2 \mcn }|\bms_k|]^{N_T/2}} \,,
}
where we have left the $\bmt$ dependence of $\bms$ implicit.
We can now compare how much data needs to be stored for this stacking procedure compared to the full data set.
Again, we have $N_T$ subintervals, each with $\Delta N$ Fourier components, and for each we have $2\mcn$ components in our data vector.
As \eqref{eq:av-data} is a real symmetric matrix, we need $\mcn (2 \mcn + 1)$ components to specify it for each $k$ value.
Thus in total, we need to store $\mcn \times (2 \mcn + 1) \times \Delta N$ entries to disk, although if $\bms_k^{-1}$ has a number of zeros (associated with experiments well within or outside the coherence length $\lc$), fewer points may be required.
This number should be contrasted with the $2 \mcn \times N = 2 \mcn \times \Delta N \times N_T$ values that would be needed in the absence of a data stacking procedure.
Thus, as long as $N_T \gg \mcn$, a significant reduction in the data set can be achieved.
For simplicity, in the main body of the paper we assume that no data stacking has been performed, though it is important to keep in mind that all results we derive may also be applied to the stacked data likelihood.
An important caveat is that care should be taken when accounting for daily modulation to make sure the data is stacked with other data taken at a similar time of day, otherwise the effect can be washed out.

Finally, we briefly demonstrate using the Asimov procedure that as long as the subintervals retain sufficient frequency resolution that the signal remains well resolved, the stacked and full likelihoods are equally sensitive.
If the signal prediction remains unchanged in each subinterval, then the averaged data set defined in \eqref{eq:av-data} has the following expected value,
\es{}{
\langle [\bar{\dd}_k]^{ij} \rangle = \frac{1}{N_T} \sum_{\ell=1}^{N_T} \langle d_{k,\ell}^i d_{k,\ell}^j \rangle = \bms^t\,.
}
It is straightforward to then evaluate the equivalent Asimov $\Theta$, and one finds a result enhanced by $N_T$, but with $T \to \Delta T$ when replacing the sum over Fourier components with an integral over speed.
For instance, the equivalent of \eqref{eq:Asimov-fv} has $T \to N_T \Delta T$.
Yet as $N_T \Delta T = T$, by definition, the test statistic is identical, and therefore the stacking procedure is optimal as claimed.

\bibliography{DMinterferometry}

\end{document}